\documentclass[fleqn,aps,prb,twocolumn,showpacs,floatfix,longbibliography]{revtex4-2}
\usepackage{amsmath,amssymb,amsfonts,float,graphics,epsfig,epstopdf,color,verbatim,tabularx,bm,multirow,appendix,hyperref}
 
\usepackage{amsthm}
\theoremstyle{definition}

\usepackage[normalem]{ulem}
\usepackage{amsmath}
\usepackage{amssymb}
\usepackage{mathtools}
\usepackage{graphicx}
\usepackage{lipsum}
\usepackage{bm}
\usepackage{hyperref}
\usepackage{url}
\usepackage[utf8]{inputenc}
\usepackage{subfigure}
\usepackage{slashed,bbm}
\usepackage{graphics,psfrag,epsfig}
\usepackage{dsfont}
\usepackage{setspace}
\usepackage{wasysym}
\usepackage{slashed}
\usepackage{lipsum}
\usepackage{braket}
\usepackage{physics}
\usepackage{enumerate}%
\usepackage{makecell}
\usepackage{xcolor}
\definecolor{dark-red}{rgb}{0.4,0.15,0.15}
\definecolor{dark-blue}{rgb}{0.15,0.15,0.4}
\definecolor{medium-blue}{rgb}{0,0,0.5}
\hypersetup{
colorlinks, linkcolor={dark-blue},
citecolor={dark-blue}, urlcolor={medium-blue}
}

\newcommand{\be}{\begin{equation}}
\newcommand{\ee}{\end{equation}}
\newcommand{\bea}{\begin{eqnarray}}
\newcommand{\eea}{\end{eqnarray}}

\begin{document}

\title{
 Non-perturbative constraints on stability and renormalization group flows in nonequilibrium matter}

\author{Yu-Hsueh Chen}
\affiliation{Department of Physics, University of California at San Diego, La Jolla, California 92093, USA}
\author{Tarun Grover}
\affiliation{Department of Physics, University of California at San Diego, La Jolla, California 92093, USA}

\begin{abstract}
We derive constraints on renormalization group (RG) flows and stability of phases in nonequilibrium systems using quantum information inequalities. These constraints involve conditional mutual information (CMI), which quantifies correlations between spatially separated regions not mediated by their surroundings. First, assuming CMI is UV finite, we derive a monotonicity constraint on its crossover scaling function. {Under certain assumptions, this implies that the CMI scaling exponent cannot increase along the RG flow.} 
Second, we bound the CMI of a convex mixture of states in terms of the CMI of individual components. We use this  inequality to infer perturbative stability of spontaneous symmetry breaking states against quantum channels that explicitly break symmetry. We illustrate these constraints through several examples, including decoherence-driven transitions in classical symmetry-broken states, area-law CMI in anisotropic conserved dynamics, and even transitions in pure quantum states. We also discuss implications for classical nonequilibrium steady states.
\end{abstract} 
\maketitle

\textcolor{red}{
}
\textbf{Introduction.} Non-perturbative constraints on renormalization group (RG) flows of interacting theories ~\cite{zamolodchikov1986irreversibility,Affleck1991universal,Friedan2004boundary,cardy1988there,komargodski2011renormalization,myers2010,Jafferis2011,casini2012,patil2025shannon} have been a powerful tool in exploring phases and phase diagrams of strongly interacting systems (see, e.g.,\cite{zamolodchikov1987renormalization,ludwig1987perturbative,Huse1984exact,appelquist1999new,anselmi1998non,luty2013theorem,klebanov2011f,grover2014entanglement}). A striking result is that such constraints can often be derived as a consequence of information theoretic inequalities~\cite{Casini04,casini2012,casini2016g,casini2019irreversibility,Casini2023entropic,harper2024g, grover2014certain}. Ideas from quantum information theory have  been used to also derive general constraints on the structure of gapped ground states and conformal field theories~\cite{Kim2013longrange,shi2020fusion,Huang2023knots,Kim2025conformal,Lin2023conformal,li2025systematic, yang2025topological}. It is natural to wonder if the nexus of ideas from information theory and many-body physics can also constrain the landscape of phenomena in more general settings, e.g., an out-of-equilibrium steady state of a long-time evolved quantum system. In this work, we will pursue this direction and employ quantum information inequalities to derive  consequences for the renormalization group flow and stability of nonequilibrium systems. We will discuss several applications, ranging from decoherence induced phase transitions to stability of spontaneous symmetry breaking against perturbations that explicitly break the symmetry, and also to classical nonequilibrium systems.

The object of our central interest will be conditional mutual information (CMI), defined as $I(A:C|B) = S(AB) + S(BC) - S(B) - S(ABC)$, where $S(X)$ denotes von Neumann entropy for the density matrix $\rho_X$ associated with region $X$. CMI is positive due to strong subadditivity (SSA) and it captures correlations between subregions $A$ and $C$ that are not mediated by $B$. In a large class of systems,  CMI decays exponentially with the separation $l_B$ between $A$ and $C$: $I \sim e^{- l_B/\xi_M}$ where $\xi_M$ is called the \textit{Markov length}. As recently discussed~\cite{sang2025stability, sang2025mixed}, Markov length plays an essential role in defining mixed state phases of matter ~\cite{coser2019classification,sang2024mixed}. Notably, Gibbs states of finite-range commuting local Hamiltonians have zero Markov length whenever $B$ shields all interaction terms connecting $A$ and $C$~\cite{clifford1971markov, leifer2008quantum,brown2012quantum}, while for more general Gibbs states of local Hamiltonians at any non-zero temperature, it is finite when $\textrm{min}(A,C) = \mathcal{O}(1)$~\cite{chen2025quantum}.

In the following, we will discuss two results. The first result is a scale-monotonicity constraint on the CMI crossover scaling function, while the second result bounds the CMI of a convex mixture of states in terms of the CMI of individual
components. 

\begin{figure}
\centering
\includegraphics[width=\linewidth]{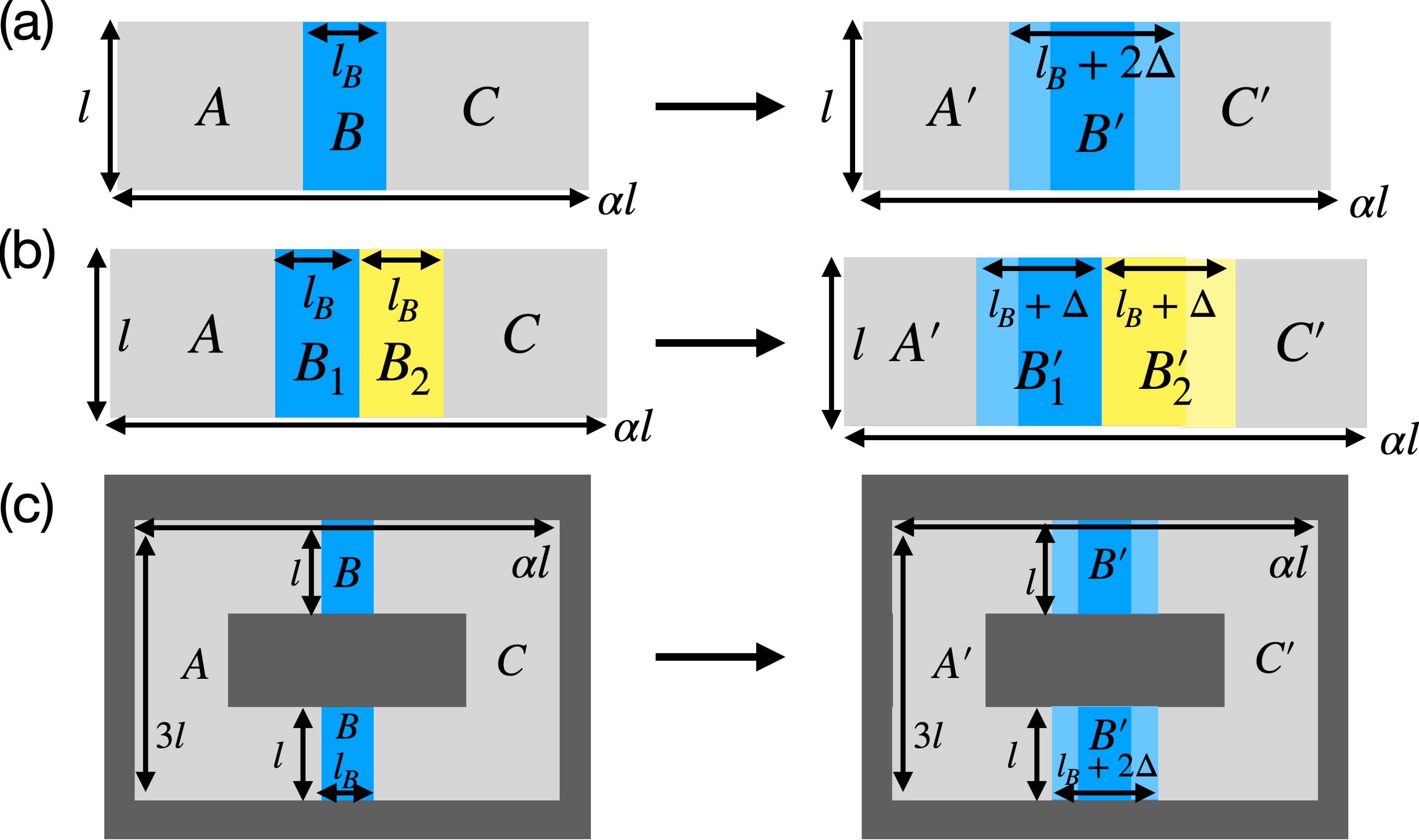}
\caption{ 
We primarily focus on the geometry shown in (a), where the full system in $d$ spatial dimensions of size $ \alpha l \times l^{d-1} $ ($\alpha$ is an arbitrary order one number) is partitioned into three regions: $A$, $B$, and $C$. $B$ is a slab of size $l_B \times l^{d-1}$ that separates $A$ and $C$ of equal sizes. (b), (c): Other geometries for which our results can be straightforwardly generalized to (see main text). 
}
\label{fig:mono_geometry}
\end{figure}

\textbf{Monotonicity constraint on CMI.} Consider a system in $d$ spatial dimensions described by a state $\rho(t)$ near a critical point, where $t$ is a dimensionless scaling variable with RG eigenvalue $1/\nu$. The full system is of size $L_\parallel \times l^{d-1}$ where $L_\parallel = \alpha l$ and $\alpha$ is any fixed O(1) number. We impose open boundary conditions along the $\parallel$ direction. We further assume that the bulk of the system is translationally invariant along $\parallel$ direction and neglect any boundary effects (we will be primarily interested in the limit $l \to \infty$). The system is partitioned into three regions, $A$, $B$, and $C$, where $B$ is a slab of size $l_B \times l^{d-1}$ that separates $A$ and $C$ which have equal sizes [see Fig.\ref{fig:mono_geometry}(a)]. $l$ and $l_B$ are both dimensionless lengths measured in units of the lattice spacing $a$, which serves as the UV cutoff. 
We will denote CMI $I(A:C|B)$ as $I(l|l_B,t)$. SSA implies that CMI is monotonically decreasing with increasing $l_B$:
\begin{equation}
\label{Eq:mono}
I(l|l_B+ 2\Delta,t) \leq I(l|l_B,t), \qquad \forall\, \Delta > 0,
\end{equation}
The proof is straightforward: the difference $\Delta I = I(A:C|B)-I(A':C'|B')$, illustrated in Fig.~\ref{fig:mono_geometry}(a), can itself be expressed as a sum of two CMIs \cite{supplement} and is therefore non-negative.

On dimensional grounds, the CMI may depend on $l_B$ and on the scaling variables
\begin{equation}
x = \operatorname{sign}(t) l_B /\xi,\quad y = l_B/l,
\end{equation}
where $\xi = 1/|t|^\nu$ is the characteristic length scale in the system.  
In the scaling limit $l_B\gg1$ with $x$ and $y$ fixed, \textit{assuming} that the CMI has a finite continuum limit (``UV finite''), one obtains $I(l|l_B,t) = f(x,y)$, and Eq.~\eqref{Eq:mono} implies
\begin{equation}
\label{Eq:mono_scale}
f(\lambda x, \lambda y) \leq f(x,y),\quad \forall \lambda \geq 1,
\end{equation}
where $\lambda$ is chosen so that the scaled geometry remains well defined. This is our first main result. 
With multiple relevant scaling parameters, the same argument constrains their joint variation along an RG scale trajectory. We restrict below to single-parameter tuned transitions. 

We next briefly discuss our assumption that the CMI is UV finite.
In a system described by a local space-time action, the UV-divergent contributions to entropies defining CMI are expected to be \textit{local functionals} of the entangling boundary~\cite{fursaev2006entanglement,ryu2006aspects,solodukhin2008entanglement,metlitski2009,Grover11_2,liu2013refinement,casini2015mutual,van2025finite}, and are therefore expected to cancel in the combination defining CMI \footnote{A prototypical example is provided by a 1+1-D CFT with central charge $c$, where the divergent part of the entanglement entropy $S_X$ for a region $X$ consisting of $N$ disjoint pieces is $2 N \times \frac{c}{6}\log(1/a)$, reflecting the additive contribution from the $2 N$ entangling boundary points~\cite{Calabrese04}. This expression remains true even when the CFT is slightly perturbed and acquires a finite correlation length $\xi \gg a$, precisely because these divergences arise from singularities localized at length scales much less than $\xi$.}. Indeed, for ground states of quantum field theories, there is substantial evidence for this statement based on general arguments~\cite{fursaev2006entanglement,ryu2006aspects,solodukhin2008entanglement,metlitski2009,Grover11_2,liu2013refinement,casini2015mutual,van2025finite} and also explicit calculations~\cite{Holzhey94,Calabrese04,levin2006detecting,vardhan2024petz,balasubramanian2019information}. 
For nonequilibrium steady states, locality of a
Martin--Siggia--Rose--Janssen--De Dominicis action ~\cite{martin1973statistical,janssen1976lagrangean,
dominicis1976techniques,tauber2014critical} similarly motivates UV finiteness of equal-time CMI.
For example, Ref.~\cite {chen2025local} analytically demonstrates that the CMI of compact directed percolation is UV finite. In the Supplemental Material ~\cite{supplement}, we provide numerical evidence for UV-finite CMI across the transition between trivial and strong-to-weak spontaneous symmetry-breaking (SWSSB) states~\mbox{\cite{lee2023quantum,lessa2024strong}}. We will provide further examples from nonequilibrium systems below that substantiate this expectation.
Together, these observations lead us to expect UV-finite CMI in systems with a
well-defined RG description.
Consistent with this expectation, known examples of non-UV-finite
CMI~\cite{Grover11_2,ma2018topological} occur in fractonic and
subsystem-symmetric systems without a conventional thermodynamic
limit~\cite{paramekanti2002ring,chamon2005glassiness,Haah2011local,
vijay2016fracton,pretko2017subdimensional,gorantla2021lowenergy,
seiberg2021exotic}. Their unconventional RG
structures~\cite{haah2014bifurcation,shirley2019foliated,dua2020bifurcating}
lie outside our scope; nevertheless, Eq.~\eqref{Eq:mono} remains valid and may
still constrain them.

We now use Eq.~\eqref{Eq:mono_scale} in the limit $l\gg \xi,l_B$, or equivalently, $y\ll1$ and $y\ll |x|$. 
This limit contains two crossover regimes: $\mathrm{UV}: 1 \gg |x| \gg y$ and $\mathrm{IR}: |x| \gg 1 \gg y$.
The ordered limits implicit in this UV/IR convention are $\mathrm{UV}:\lim_{x\to0}\lim_{y\to0^+}$ and $\mathrm{IR}:\lim_{|x|\to\infty}\lim_{y\to0^+}$ on each fixed-sign branch of $x$. 
We note that one can also define the  reversed order-of-limits conventions $\mathrm{UV}^{\prime}:\lim_{y\to0^+}\lim_{x\to0}$ and $\mathrm{IR}^{\prime}:\lim_{y\to0^+}\lim_{|x|\to\infty}$. 
These reversed limits need not agree with the UV and IR limits. We will see below that Illustration 1 has $\mathrm{UV} \neq \mathrm{UV}^{\prime}$, while Illustration 2 and 3 has $\mathrm{UV}=\mathrm{UV}^{\prime}$.

We first consider the case when $f_0(x)\equiv \lim_{y\to0^+}f(x,y)$ is finite for fixed nonzero $x$, then Eq.~\eqref{Eq:mono_scale} gives $f_0(\lambda x)\le f_0(x)$ for all $\lambda\ge1$. This implies
\begin{equation}
\label{Eq:dfdx_}
f_0(\textrm{UV}) \geq f_0(\textrm{IR}).
\end{equation}

Let's next consider the case when $f(x,y \ll 1)$ diverges as some negative power of $y$. In this case, SSA still constrains the corresponding leading power-law exponents. Suppose there exists some fixed $\lambda > 1$ so that $x_{\mathrm{IR}}=\lambda x_{\mathrm{UV}}$ with $0<|x_{\mathrm{UV}}|\ll1$ and $|x_{\mathrm{IR}}|\gg1$, while the following scaling holds:  $f(x_{\mathrm{UV}},y)=y^{-\eta_{\mathrm{UV}}+o(1)}$ and $f(x_{\mathrm{IR}},y)=y^{-\eta_{\mathrm{IR}}+o(1)}$ as $y\to0^+$. 
Eq.~\eqref{Eq:mono_scale} gives $f(x_{\mathrm{IR}},\lambda y)\le f(x_{\mathrm{UV}},y)$. Since $\lambda$ is being held fixed, this compares $y^{-\eta_{\mathrm{IR}}+o(1)}$ with $y^{-\eta_{\mathrm{UV}}+o(1)}$, and hence $\eta_{\mathrm{UV}}\ge \eta_{\mathrm{IR}}$.  {Thus the CMI in the IR cannot diverge faster than that in the UV.}

{ 
When $\eta_{\mathrm{UV}}=\eta_{\mathrm{IR}}\equiv\eta$, one can further constrain the coefficient given the following stronger asymptotic assumptions. Suppose first that $f(x,y)=g(x,y)/y^\eta$, with $\eta\ge0$ and $g_0(x)\equiv \lim_{y\to0^+}g(x,y)$ finite for fixed nonzero $x$.  Eq.~\eqref{Eq:mono_scale} then gives $g_0(\lambda x)\le \lambda^\eta g_0(x)$ for all $\lambda >1$, or
\begin{equation}
\label{Eq:dgdx_}
\frac{g_0(x_{\text{UV}})}{|x_{\text{UV}}|^\eta}
\geq  \frac{g_0(x_{\text{IR}})}{|x_{\text{IR}}|^\eta},
\end{equation}
Therefore, $g_0(x)/|x|^\eta$ is monotonically decreasing (we note that the $\eta =0$ case reduces to Eq.~\eqref{Eq:dfdx_}). 
For CMI that is extensive in the transverse area, one generally expects $\eta=d-1$, where $d$ is the spatial dimension.  
Below we will discuss an application of Eq.\eqref{Eq:dgdx_} to a $(d+1)$-D non-equilibrium system.
}

We now discuss implications when $f_0(x)$ exists and assume $f_0(\mathrm{UV})=\lim_{y\to0^+}f(0,y)$.
Consider perturbing a Gibbs state of a finite-range commuting local Hamiltonian
by an arbitrary local channel of strength $p$, with $B$ shielding all
interactions between $A$ and $C$. Even if the unperturbed state is a finite-$T$
critical point with relevant RG directions, its CMI vanishes exactly; hence,
monotonicity forbids destabilization toward a state with nonzero
CMI~\cite{clifford1971markov,leifer2008quantum,brown2012quantum}. The same
conclusion was recently derived microscopically for specific
channels~\cite{zhang2025stability}. As a consistency check, a relevant
nonlinear term that breaks detailed balance drives a Gaussian Gibbs state to
the $1+1$-D KPZ fixed point~\cite{kardar1986dynamic}.
Under these assumptions, monotonicity gives zero scaling-limit CMI for the KPZ
steady state, consistent with its representation as a Gibbs state of a local
Hamiltonian~\cite{kardar1986dynamic}. 

{More generally, the same monotonicity rules out RG flows to a fixed point with
larger CMI. For the full-system partition in Fig.~\ref{fig:mono_geometry}(a),
where $A\cup B\cup C$ covers the system and the global charge constraint remains
visible, an infinitesimal perturbation can drive a $\mathbb{Z}_2$
strong-to-weak spontaneously symmetry-breaking (SWSSB) fixed point with CMI
$\log 2$ to a trivial state with zero CMI, but not to a $\mathbb{Z}_4$ SWSSB
fixed point with CMI $\log 4$~\cite{lee2023quantum,lessa2024strong}. We briefly
review SWSSB in Ref.~\cite{supplement}.

}

Conversely, an IR fixed point with infinite Markov length requires the UV fixed
point to have infinite Markov length and thus rules out its representation as a
Gibbs state of a finite-range commuting local Hamiltonian. This constrains the
exotic multicritical point of Ref.~\cite{ginelli2003multiplicative}, which flows
to the $1+1$-D directed-percolation (DP) critical point on one side of the phase
diagram and to the $1+1$-D KPZ critical point on the other. Because the DP
critical point has infinite Markov length~\cite{chen2025local},\footnote{For a 1+1-D system, the CMI in our geometry is determined by the entropy $S(r)$ corresponding to the marginal probability distribution for a subsystem of size $r$. Ref.\cite{chen2025local} studied CMI in a geometry which extracts $\partial^2 S/\partial r^2$, and found that this second derivative scales as $1/r^2$. This implies that for our geometry Fig.\ref{fig:mono_geometry} , CMI is logarthimically divergent in $y = \ell/\ell_B$} the
multicritical state must as well and cannot be such a Gibbs state, although its
critical field theory remains unknown.

Although we focus on Fig.~\ref{fig:mono_geometry}(a), the results extend to
other geometries. In panels (b) and (c), respectively, a $1+1$-D
$\mathbb{Z}_2\times\mathbb{Z}_2$ symmetry-protected topological (SPT) state and
$2+1$-D topologically ordered states have nonzero CMI.
For panel (b), the chain rule and SSA show that equally enlarging $B_1$ and the
traced-out region $B_2$ cannot increase CMI~\cite{supplement}.
The same scaling argument yields Eq.~\eqref{Eq:mono_scale} for this geometry,
and panel (c) follows analogously.

\textbf{Bounding CMI via convex decomposition.} Next, we will employ SSA to derive a bound on CMI for a convex sum of density matrices in terms of the CMI of individual components. Consider any convex decomposition $\rho = \sum_{\alpha} p_{\alpha} \rho_{\alpha }$. Using SSA, one finds \cite{supplement}:
\begin{equation}
\label{Eq:nice_inequality}
\begin{aligned}
I_\rho(A:C|B) \leq  &  \sum_{\alpha} p_{\alpha} I_{\rho_{\alpha}}(A:C|B)  + S_\sigma(D|B),
\end{aligned}
\end{equation}
where $\sigma = \sum_\alpha p_\alpha |\alpha\rangle \langle \alpha|_D \otimes \rho_{\alpha}$. Eq.\eqref{Eq:nice_inequality} can be thought of as the analog Holevo bound, $S(\rho) \leq H(\{ p_\alpha\}) + \sum_\alpha p_\alpha S(\rho_\alpha)$, for CMI. %

Eq.~\eqref{Eq:nice_inequality} implies that if each $\rho_\alpha$ has finite Markov length and $S_\sigma(D|B)$ decays exponentially in $l_B$, then $\rho$ also has a finite Markov length. This observation allows one to assess robustness of spontaneous symmetry-breaking against generic perturbations, including those that break the symmetry explicitly. As a concrete example, consider the classical mixture $\rho_0 = \tfrac{1}{2} (|\uparrow\rangle\langle\uparrow|)^N + \tfrac{1}{2}(|\downarrow\rangle\langle\downarrow|)^N$ {which describes the zero-temperature limit of the classical Ising model in any dimension and has long range $\langle Z_i Z_j \rangle $ correlations}. 
Applying a $p$-bounded local channel $\mathcal{E}$ of strength $p$ \footnote{Here $p$-bounded noise means that, upon measuring $\mathcal{E}\big[(|\uparrow\rangle\langle\uparrow|)^{\otimes L}\big]$ in the computational basis, the marginal probability of observing an all-down configuration on any subregion $X$ is bounded by $p^{|X|}$, where $|X|$ denotes the size of $X$. An analogous statement holds for $\mathcal{E}\big[(|\downarrow\rangle\langle\downarrow|)^{\otimes L}\big]$.
} to $\rho_0$ leads to a natural decomposition,
\begin{equation}
\label{Eq:SSB_stable}
\rho = \tfrac{1}{2}\,\mathcal{E}\big[(|\uparrow\rangle\langle\uparrow|)^{\otimes N}\big] + \tfrac{1}{2}\,\mathcal{E}\big[(|\downarrow\rangle\langle\downarrow|)^{\otimes N}\big].
\end{equation}
Under the assumption that a trivial product state (e.g., $(|\uparrow\rangle\langle\uparrow|)^N$) remains robust under the action of $\mathcal{E}$, the first term in Eq.~\eqref{Eq:nice_inequality} decays exponentially with $l_B$ {for sufficiently small $p$}. This assumption is equivalent to requiring that the trivial phase corresponding to a product state is a stable phase of matter, {which has been proven in Ref.~\cite{zhang2025conditional} for a broad class of channels}.   
Moreover, Fano's inequality \cite{fano1961transmission} implies that $S_\sigma(D|B)$ also decays exponentially for any $p$-bounded channel with sufficiently small $p$. 
Therfore, the Markov length remains finite throughout any local path satisying $0\leq p'\leq p$. Within the local-reversibility framework, the decohered density matrix is therefore in the same phase as $\rho_0$~\cite{sang2025stability,sang2025mixed} and it guarantees a reverse quasi-local channel $\mathcal{E}^{-1}$ such that $\mathcal{E}^{-1}[\rho(p)]=\rho_0$~\cite{sang2025stability,sang2025mixed}. As an aside, let $T$ be any $\mathbb{Z}_2$-symmetric channel whose steady state is $\rho_0$. The existence of $\mathcal{E}^{-1}$ then allows one to construct a quasi-local channel $\mathcal{E}T\mathcal{E}^{-1}$ whose steady state is $\rho(p)$---\textit{a state with long-range order, but without Ising symmetry}, i.e., $\prod_i X_i \,\rho(p)\, \prod_i X_i \neq \rho(p)$~\footnote{We thank John McGreevy for a helpful discussion on this point}. This is reminiscent of Toom's model~\cite{toom1974nonergodic,toom1980stable}.
We  briefly comment on generalization to other binary classical codes in Ref.~\cite{supplement}.

We will now consider a few additional applications/demonstrations of our general results.

\textbf{Illustration 1: Stability of classical SSB and its RG flow.} We study a simple model illustrating both Eq.\eqref{Eq:dfdx_} and Eq.\eqref{Eq:nice_inequality}.
We subject the state $\rho_0 = \tfrac{1}{2} (|\uparrow\rangle\langle\uparrow|)^l + \tfrac{1}{2}(|\downarrow\rangle\langle\downarrow|)^l$ in $d=1$ to a channel $\mathcal{E}(p)$ that independently flips each $\downarrow$ spin with probability $p$ while leaving  $\uparrow$ spins unchanged. This produces a one-parameter family of mixed states 
$\rho(p) = \frac{1}{2}(|\uparrow \rangle \langle \uparrow|)^{ l} + \frac{1}{2}[(1-p)|\downarrow \rangle \langle \downarrow|+p|\uparrow \rangle \langle \uparrow|]^{ l}$. We first note that $\rho(p)$ is not a finite-temperature Gibbs state of any finite-range local Hamiltonian: for every finite block $\Lambda$, $\Pr_{\rho(p)}(\uparrow^\Lambda)=\tfrac12(1+p^{|\Lambda|})\geq\tfrac12$, so this probability remains bounded below by $1/2$ for arbitrarily large blocks, violating the alignment-suppression property of finite-temperature Gibbs states~\cite{fernandez2001non, lebowitz1988pseudo}. 
 
\begin{figure}
\centering
\includegraphics[width=0.85\linewidth]{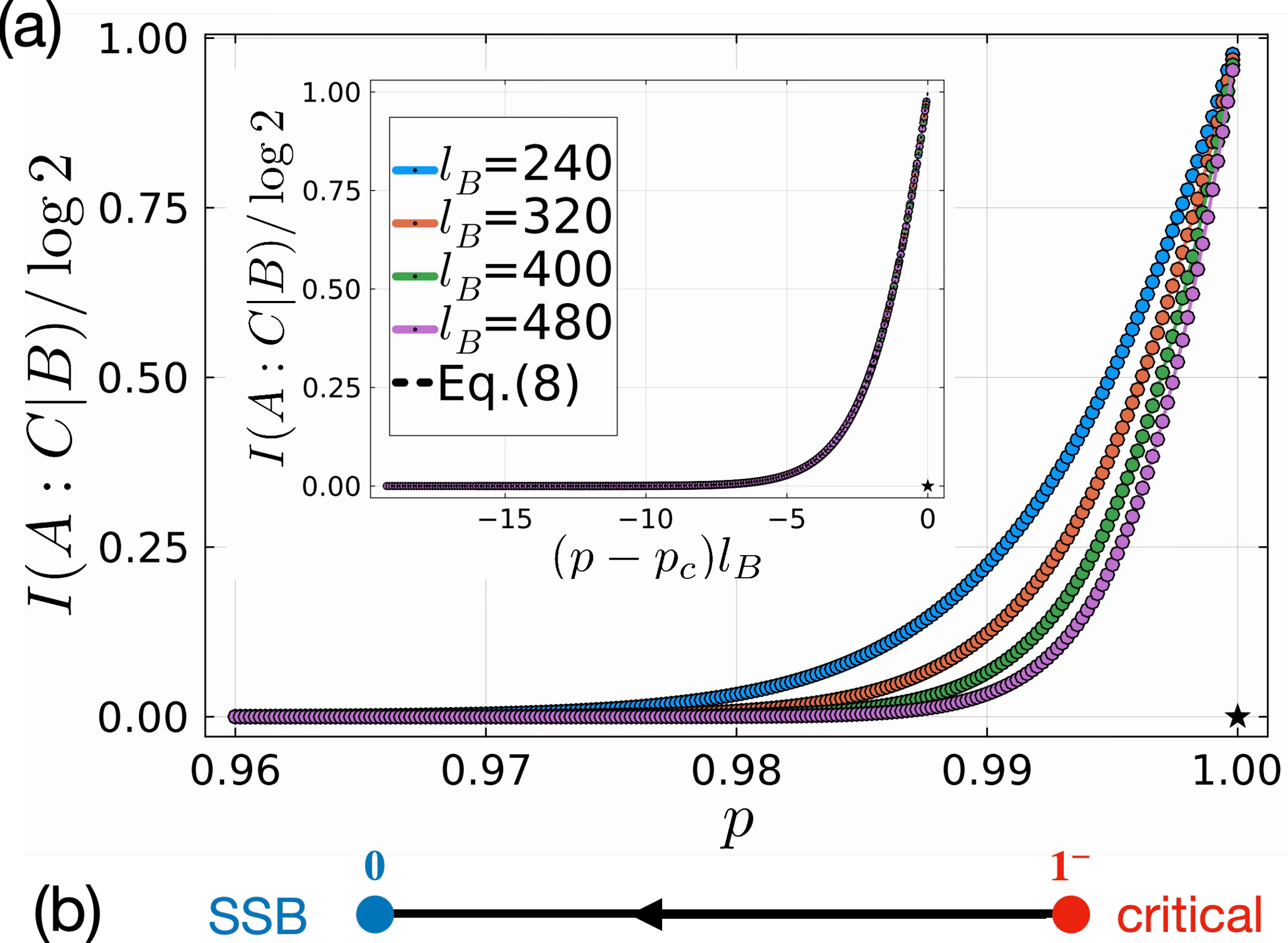}
\caption{ (a) The conditional mutual information(CMI) $I(A:C|B)=I(l|l_B,p)$ of the mixed state discussed in Illustration 1 as a function of $p$ for various $l_B$ with $ l = 10^8 (\approx \infty)$. The inset shows the data collapse of the scaling ansatz $I(l|l_B,p)=f[(p-p_c)l_B^{1/\nu}]$ with $p_c = 1$ and $\nu = 1$. (b) The corresponding RG flow.
}
\label{fig:cmi_max_bias}
\end{figure}

Let us use Eq.~\eqref{Eq:nice_inequality} to establish finite Markov length throughout this path for sufficiently small $p$. Since both $\mathcal{E}[(|\uparrow\rangle\langle\uparrow|)^{ l}]$ and $\mathcal{E}[(|\downarrow\rangle\langle\downarrow|)^{ l}]$ are product states, the first term on the right-hand side of Eq.~\eqref{Eq:nice_inequality} vanishes. Consequently, $I_\rho(A:C|B)$ decays exponentially whenever $S_{\sigma}(D|B)$ does, which holds for any $p$-bounded noise with sufficiently small $p$ \footnote{The existence of a nonzero threshold shows that violating alignment suppression does not imply that $\rho(p)$ is distinct from the classical SSB phase, and hence is not a universal characteristic of non-Gibbsianness}. We next compute the CMI of $\rho$ explicitly in the limit $l\to\infty$ to illustrate that the associated scaling function is monotonic. A direct calculation yields
$I(\infty|l_B,p)=\tfrac{1}{2}(1+p^{l_B})\log(1+p^{l_B})-\tfrac{1}{2}p^{l_B}\log(p^{l_B})$.
In the limit $\epsilon\equiv(p-1)\to 0^-$ and $l_B\to\infty$ with $x=\epsilon l_B$ held fixed, one finds
\begin{equation}
\label{Eq:illus_cmi}
I(\infty|l_B,p)\approx f(x)\equiv \tfrac{1}{2}\big[(1+e^{x})\log(1+e^{x})-x e^{x}\big].
\end{equation}
This identifies $p_c=1$ and the correlation-length exponent $\nu=1$, and more importantly shows that $I(\infty|l_B,p)$ is UV finite. The scaling function $f(x)$ is readily verified to be monotonically increasing for $x\in(-\infty,0)$ [see Fig.~\ref{fig:cmi_max_bias}(a)]. 

We now analyze $f(x)$ in the critical and stable regimes. When $x$ approaches $0^{-}$, one finds $I(\infty|l_B,p)\approx \log 2\,(1+x/2)$, while for $x \ll 0$, $I(\infty|l_B,p)\approx e^{x}(1-x)/2$, which  vanishes exponentially as $x\to -\infty$; see Fig.~\ref{fig:cmi_max_bias}(a).   
Taking $x\to0^-$ within the scaling regime gives $f(x)\to\log 2$, whereas setting $p=1$ before taking the scaling limit yields an exact product state with zero CMI. Thus, the UV and UV' fixed points do not coincide, as noted above. 
This is expected because $\rho(p=1)$ is a product state and cannot be destabilized towards a symmetry-broken state.

Finally, for $l = l_B + 2r$ with finite $r$, we find $I(l_B + 2r|l_B,p)=(r/l_B)^{\eta} f(x)$ with anomalous dimension $\eta=2>0$ \cite{supplement}, consistent with the aforementioned SSA constraint that the CMI is monotonically increasing under increasing $l$ while keeping $l_B$ fixed. This finite-$r$ regime is distinct from the RG scaling limit $l/l_B\to\infty$ considered above. We also consider a more general channel that flips a $\downarrow$($\uparrow$) spin with probability $p$($q$) \cite{supplement}; the case $p=q$ was also studied previously in Ref.\cite{lloyd2025diverging}. While most features persist from the $q=0$ case, we find that the correlation-length exponent is $\nu=2$ for fixed $0<q<1$, away from the crossover to the $q=0$ regime, in contrast to $\nu=1$ at $q=0$. 

\begin{figure}
\centering
\includegraphics[width=\linewidth]{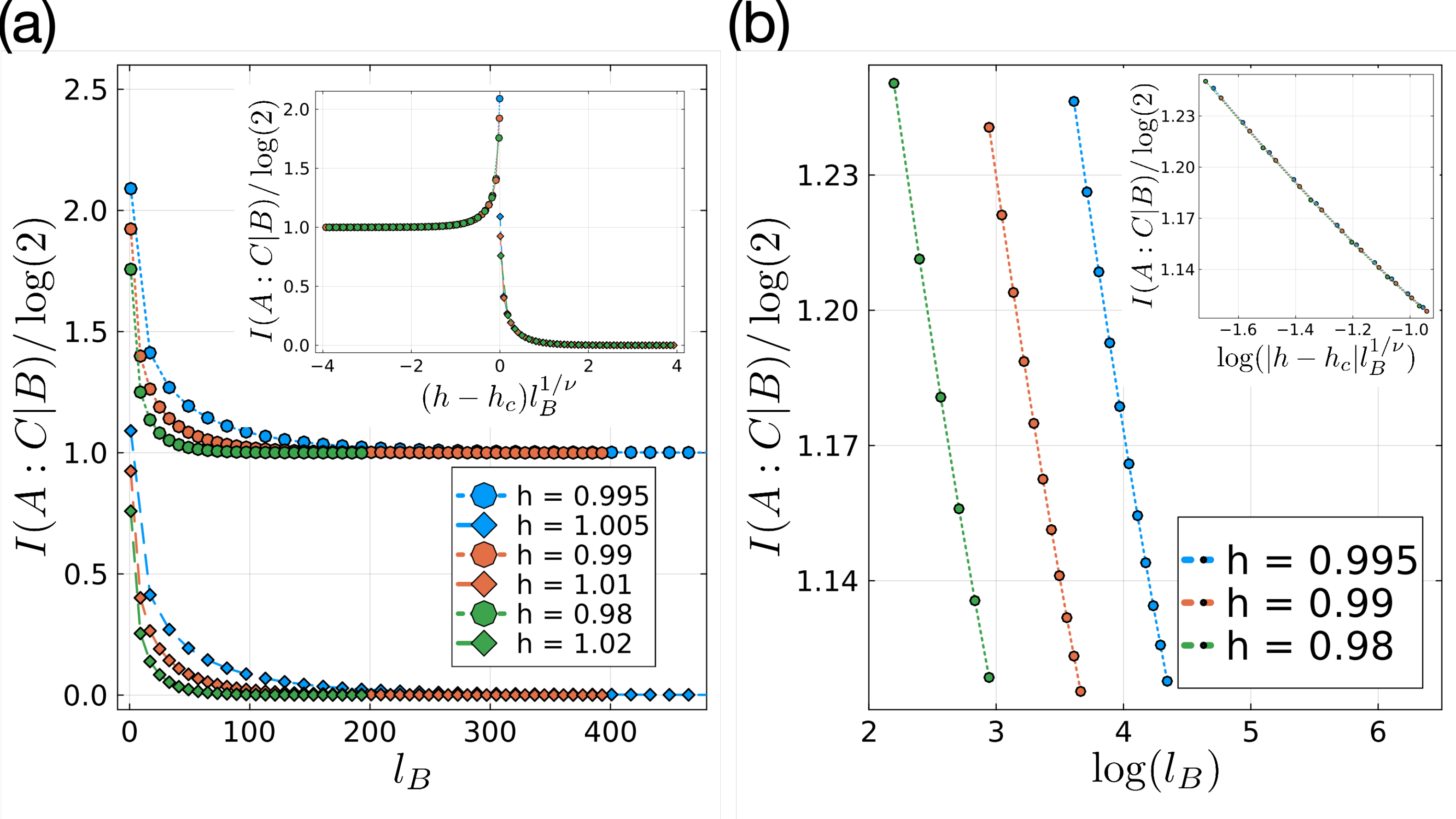}
\caption{ 
(a) The CMI of the TFIM as a function of $l_B$ for several values of $h$. 
The inset shows a data collapse using the scaling ansatz $I(A:C|B) = f[(h-h_c)l_B^{1/\nu}]$ with $h_c = 1$ and $\nu = 1$. 
(b) The CMI as a function of $\log l_B$ for several values of $h < h_c$. 
The insets show the same data plotted as a function of $\log(|h-h_c|\,l_B^{1/\nu})$.
}
\label{fig:tfim}
\end{figure} 

\textbf{Illustration 2: CMI of TFIM.} The above constraints apply also to pure states. Let's consider $ABC$ spanning the entire system so that the CMI reduces to the mutual information $I(A:C)$. As an illustration, we compute the CMI of the ground state of the $(1+1)$D transverse-field Ising model (TFIM) with Hamiltonian $H=-\sum_j Z_j Z_{j+1}-h\sum_j X_j$, which is efficiently accessible by mapping to free fermions. 
For each finite $l$, we use the unique $\mathbb{Z}_2$-even ground state, so the result in the ordered phase retains the cat-like contribution. Taking the thermodynamic limit before removing an infinitesimal symmetry-breaking field instead selects a symmetry-broken state with a different $O(1)$ contribution, which we do not compare here.  
We choose the system size as $l =12/|h-h_c|^{\nu}+1$, with $\nu=1$ and $h_c=1$, which is sufficiently large that the CMI is well described by the scaling form, even though strict validity requires $l\to\infty$.

Fig.~\ref{fig:tfim}(a) shows the CMI as a function of $l_B$ for several values of $h$, while the inset plots it as a function of $x=(h-h_c)l_B$, demonstrating the scaling form $I(A:C|B)=f(x)$. One observes that $f(x)$ is monotonically decreasing (increasing) for $x\in(0,\infty)$ ($x\in(-\infty,0)$). Further, when $x\to 0^-$, $f(x)\approx -(c/3)\log(|x|)+U_{\text{ising}}+\log 2$ diverges logarithmically with central charge $c=1/2$---the constant subleading term is well-defined since $f(x)$ is independent of the UV cutoff (here $U_{\text{ising}} \approx -0.131984$). This behavior follows analytically from Ref.~\cite{calabrese2004entanglement, cardy2008form}, and the logarithmic divergence is confirmed clearly in Fig.~\ref{fig:tfim}(b). A similar analysis yields $f(x)\approx -(c/3)\log (|x|) + U_{\text{ising}}$, again diverging logarithmically as $x\to 0^+$. 
{This divergence implies that, unlike the SSA construction of Ref.\cite{Casini04} that results in c-theorem, no nontrivial constraint arises for RG flows between 1+1-D CFTs.}

{
\textbf{Illustration 3: SSA constraint on area-law CMI in anisotropic conserved dynamics.}
Finally, we illustrate Eq.~\eqref{Eq:dgdx_} for Gaussian conserved dynamics introduced in Ref.~\cite{grinstein1990conservation}
in the context of self-organized criticality in $d>1$. In this example, CMI satisfies an ``area-law'': the scaling function $f(x,y) \sim y^{-(d-1)} = (l/l_B)^{d-1}$ both in the UV and the IR, i.e., $\eta_{\text{UV}} = \eta_{\text{IR}} = d-1$, as detailed in Ref.~\cite{supplement}. Factoring out $(l/l_B)^{d-1}$ leaves the dimensionless coefficient $g_0(x)$, which is constrained by
Eq.~\eqref{Eq:dgdx_} even though the total CMI is not $O(1)$.

Here a coarse-grained conserved field obeys the Langevin equation
$\partial_t\phi=\mu\nabla^2\phi-\nabla^4\phi+\zeta$, 
where the zero-mean conserved noise has covariance
$\langle\zeta(\mathbf q,t)\zeta(\mathbf q',t')\rangle
=2(D_\parallel q_\parallel^2+D_\perp q_\perp^2)(2\pi)^d
\delta^d(\mathbf q+\mathbf q')\delta(t-t')$. 
Defining $\delta\equiv1-D_\perp/D_\parallel$, the
nonzero-mode steady-state measure is local at $\delta=0$, and its area-law CMI
therefore vanishes exactly. {Weak anisotropy instead produces a nonlocal
steady-state measure~\cite{grinstein1990conservation}.}
Explicitly, the steady-state measure is
$P_{\mathrm{ss}}[\phi]\propto\exp[-\frac{1}{2}\int_{\mathbf q\neq0}
\phi(\mathbf q)K(\mathbf q)\phi(-\mathbf q)]$, where
$\int_{\mathbf q}\equiv\int d^d q/(2\pi)^d$.
With $q_\parallel$ and $q_\perp$ denoting momenta normal and tangent to the
interfaces, respectively, its precision kernel on the nonzero-momentum
subspace decomposes exactly as $K=K_{\mathrm{loc}}+K_{\mathrm{nloc}}$, where
$K_{\mathrm{loc}}=[\mu+q_\parallel^2+(1+\delta)q_\perp^2]/D_\parallel$ is
polynomial in momentum and hence local in real space, while 
\begin{equation}
\begin{aligned}
K_{\mathrm{nloc}}
=\frac{\delta\mu q_\perp^2+\delta^2q_\perp^4}
{D_\parallel\left[q_\parallel^2+(1-\delta)q_\perp^2\right]},
\end{aligned} 
\label{Eq:soc_precision_decomposition}
\end{equation}
leads to non-locality, see Ref.~\cite{supplement} for details.
Under $\mathbf q\to\mathbf q/b$, $\delta$ appears only in dimensionless ratios
of momentum components and is therefore marginal, whereas
$\mu\to b^2\mu$ is relevant and generates a diverging scale $\xi=\mu^{-1/2}$. We now treat
$\delta$ as an independent small parameter and compare $g_0(x;\delta)$ in the
UV ($x\ll1$) and IR ($x\gg1$), where $x=l_B/\xi$.
Crucially,  Eq.~\eqref{Eq:soc_precision_decomposition} shows that the nonlocal kernel
starts at $O(\delta^2)$ at the UV fixed point ($\mu=0$), but at $O(\delta)$ in
the IR, where the $q^4$ term is irrelevant.
 
{
We now translate this kernel scaling into CMI scaling. 
When the nonlocal term
vanishes (i.e., $\delta = 0$), the CMI is zero.
Since  CMI is nonnegative, its leading dependence generically scales quadratically with the nonlocal term. 
Eq.~\eqref{Eq:soc_precision_decomposition} then implies generically, $g_0(0^+;\delta)=c_0\delta^4 $ and $g_0(\infty;\delta)=c_\infty \delta^2$ with $c_0, c_\infty > 0$. One might naively expect that when $x \ll 1$, the leading term in $g_0(x)$ will still scale as $\delta^4$. 
We now show that, given these endpoint values, this expectation is not true when $x^{d-1} \gg \delta^2$. In particular, SSA puts a nontrivial constraint on $g(x;\delta)$ when $x \ll 1$, \textit{assuming} there exists some fixed $x_{\text{IR}} \gg 1$ independent of $\delta$ such that $g_0(x_{\text{IR}};\delta)=c_{x_{\text{IR}}} \delta^2 +o(\delta^2),\ c_{x_{\text{IR}}} >0$ holds.
Eq.~\eqref{Eq:dgdx_} requires
$g_0(x;\delta)/x^{d-1}$ to be monotonically decreasing.  
Comparing  UV and IR fixed points then gives
$g_0(x;\delta) \gtrsim \delta^2{x^{d-1}}$, when $ x \ll 1$ for sufficiently small $\delta$.
We emphasize that this constraint requires no explicit evaluation of
the CMI: it follows simply from its quadratic onset, the kernel power counting, and 
SSA.  

{
In Ref.~\cite{supplement}, we explicitly verify that when $d =2$, the steady state of the anisotropic conserved dynamics \textit{saturates} the aforementioned bound; specifically,
$g_0(x;\delta)=\delta^2x/(120\pi) +o(x)$ for $\delta^2\ll x\ll1$. }

}

We have shown that SSA constrains the scaling of CMI along RG flows,
including regimes in which the CMI diverges as a power of $l/l_B$, while
our convex-decomposition bound provides stability criteria for symmetry-broken
states under local channels. Together, these results provide a nonperturbative
information-theoretic framework for organizing nonequilibrium fixed points
and phases.
It will be interesting to explore our CMI constraints for strongly interacting non-equilibrium systems in $d > 1$ such as flocking fixed points and their RG flows~\cite{toner1995long,chate2020dry,toner2024physics}.

\textbf{Acknowledgments:}
We thank John McGreevy for helpful discussions, Shengqi Sang and Tim Hsieh for helpful feedback on the manuscript, and Mehran Kardar for pointing out Ref.\cite{grinstein1990conservation}. T.G. is supported by the National Science Foundation under Grant No. DMR-2521369. We acknowledge the hospitality of
Kavli Institute for Theoretical Physics (KITP) and thank
the organizers of the KITP programs ``Noise-robust Phases of Quantum Matter'', and ``Learning the Fine Structure of Quantum Dynamics in Programmable Quantum Matter''. This research was supported in part by grant NSF PHY-2309135
to the KITP.  

\bibliography{bibs}

\onecolumngrid 

\clearpage %

\appendix
 
\section{Proofs of CMI constraints using SSA}

\subsection*{Proof of Eq.(1) in the main text}

We now show that $I(l|l_B)\equiv I(A:C|B)\geq I(l|l_B+2\Delta)\equiv I(A':C'|B'')$, for all $\Delta>0$ [see Fig.~\ref{fig:supp_mono_cmi}(a) and (c)], which corresponds to Eq.~(1) in the main text. It is conceptually convenient to introduce an intermediate configuration $A B'C'$ [Fig.~\ref{fig:supp_mono_cmi}(b)] that interpolates between $ABC$ and $A'B''C'$. In particular, we establish the chain of inequalities $I(A:C|B)\geq I(A:C'|B')\geq I(A':C'|B'')$.

Let us define $\delta\equiv B'\setminus B$, so that $B'=B\delta$ and $C'=C\setminus\delta$. The key observation underlying $I(A:C|B)\geq I(A:C'|B')$ is that $B'C'=BC$, which implies $I(A:C'|B')=S(AB')+S(B'C')-S(B')-S(AB'C')=S(AB\delta)+S(BC)-S(B\delta)-S(ABC)$. Therefore,
\begin{equation}
\begin{aligned}
I(A:C'|B')-I(A:C|B)
&=[S(AB\delta)+S(BC)-S(B\delta)-S(ABC)]-[S(AB)+S(BC)-S(B)-S(ABC)] \\
&=-[S(AB)+S(B\delta)-S(B)-S(AB\delta)] \\
&=-I(A:\delta|B)\leq 0 ,
\end{aligned}
\end{equation}
where the final inequality follows from SSA. The same calculation shows that $I(A:C'|B')\geq I(A':C'|B'')$, and hence $I(A:C|B)\geq I(A':C'|B'')$.

\begin{figure}
\centering
\includegraphics[width=\linewidth]{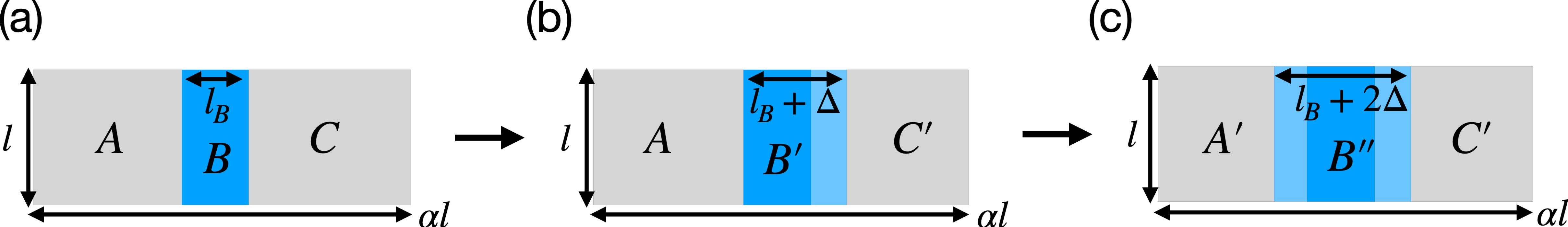}
\caption{ To prove that $I(A:C|B)\geq I(A':C'|B'')$, we first introduce an intermediate configuration $AB'C'$ in (b) that interpolates between $ABC$ in (a) and $A'B''C'$ in (c), and then establish the chain of inequalities $I(A:C|B)\geq I(A:C'|B')\geq I(A':C'|B'')$.
}
\label{fig:supp_mono_cmi}
\end{figure}

\subsection*{Four-region geometry in Fig.~1(b)}

We now prove the monotonicity statement for the four-region geometry in Fig.~1(b) of the main text. Let $X\subset A$ and $Y\subset C$ be strips of width $\Delta$ adjacent to $B_1$ and $B_2$, respectively, and write
$A=A'X$, $C=YC'$, $B_1'=XB_1$, and $B_2'=B_2Y$. The deformation therefore enlarges both $B_1$ and the unconditioned region $B_2$ by $\Delta$. All CMIs below are evaluated after tracing out $B_2$ in the initial partition or $B_2'$ in the final partition. Applying the chain rule twice gives
\begin{equation}
\begin{aligned}
I(A:C|B_1)-I(A':C'|B_1')
&=I(A'X:YC'|B_1)-I(A':C'|B_1X)\\
&=I(A'X:Y|B_1C')+I(X:C'|B_1)\geq0 .
\end{aligned}
\end{equation}
Since both terms in the final line are nonnegative by SSA, this proves that the CMI cannot increase under the deformation shown in Fig.~1(b).

\subsection*{Proof of Eq.(6) in the main text}

It is illuminating to first rewrite Eq.~(3) in the main text solely in terms of the classical--quantum state $\sigma = \sum_\alpha p_\alpha |\alpha \rangle \langle \alpha |_D \otimes \rho_\alpha$. Clearly, $I_\sigma(A:C|B) = I_\rho(A:C|B)$ and $\sum_{\alpha} p_\alpha I_{\rho_\alpha}(A:C|B) = I_\sigma(A:C|B,D)$. 
Eq.~(3) is therefore equivalent to 
\begin{equation}
\label{Eq:equivalent}
I_\sigma(A:C|B) \leq I_\sigma(A:C|B,D) +  S_\sigma(D|B).
\end{equation}
Since all quantities are now expressed in terms of $\sigma$, we will omit the subscript in $I$ and $S$ in the following for notational simplicity.

A straightforward use of the definition of CMI shows that
\begin{equation}
\label{Eq:supp_intermediate}
\begin{aligned}
I(A:C|B) - I(A:C|B,D) & = \Big[ S{(AB)} + S{(BC)} - S(B) - S{(ABC)}\Big] - \Big[ S{(ABD)} + S{(BCD)} - S{(BD)} - S{(ABCD)}\Big] \\
& = \Big[ ( S{(BD)} - S(B)) -(S{(ABD)}  - S{(AB)})  \Big] - \Big[ S{(ABC)} + S{(BCD)} - S{(BC)} - S{(ABCD)}\Big]  \\
& = \Big[ S(D|B)  - S(D|AB) \Big]  - I(A:D|BC)\\
&  \leq   S(D|B)  - S(D|AB),
\end{aligned}
\end{equation}
where we have use the strong subadditivity in the final line. Eq.\eqref{Eq:supp_intermediate} is fairly general for any quadripartition $A, B,C,D$. Now, using the fact that $\sigma = \sum_\alpha p_\alpha |\alpha \rangle \langle \alpha |_D \otimes \rho_\alpha$ and thus $D$ is only classicaly correlated with the rest of the system, one has $S(D|X) \geq 0$ for any $X \in ABC$.  Choosing $X = AB$ , one then has $S(D|B)  - S(D|AB) \leq S(D|B) $, and thus the desired inequality Eq.\eqref{Eq:equivalent} holds.

\section{Proof that $S(D|B)$ decays exponentially with $N_B$}

As discussed in the main text, the CMI of $\rho$ can be bounded using a convex decomposition $\rho=\sum_\alpha p_\alpha \rho_\alpha$ as
$I_\rho(A:C|B)\leq \sum_\alpha p_\alpha I_{\rho_\alpha}(A:C|B)+S_\sigma(D|B)$,
where $\sigma=\sum_\alpha p_\alpha |\alpha\rangle\langle\alpha|_D\otimes\rho_\alpha$.
Therefore, showing that both $S_\sigma(D|B)$ and each $I_{\rho_\alpha}(A:C|B)$ decay exponentially with $l_B$ is sufficient to establish that $I_\rho(A:C|B)$ decays exponentially. For the classical SSB discussed in the main text, $\rho = \tfrac{1}{2}\,\mathcal{E}\big[(|\uparrow\rangle\langle\uparrow|)^{\otimes N}\big] + \tfrac{1}{2}\,\mathcal{E}\big[(|\downarrow\rangle\langle\downarrow|)^{\otimes N}\big]$ and thus
\begin{equation}
\sigma = \tfrac{1}{2}|\alpha = \uparrow\rangle \langle \alpha = \uparrow|_D \otimes \mathcal{E}\big[(|\uparrow\rangle\langle\uparrow|)^{\otimes N}\big] + \tfrac{1}{2}|\alpha = \downarrow\rangle \langle \alpha = \downarrow|_D \otimes \mathcal{E}\big[(|\downarrow\rangle\langle\downarrow|)^{\otimes N}\big].
\end{equation}
The goal of this appendix is to show that for any $p$-bounded channel there exists a threshold $p_c$ such that, for all $p<p_c$, $S_\sigma(D|B)$ decays exponentially as a function of $l_B$.
Here $p$-bounded noise means that, upon measuring $\mathcal{E}\big[(|\uparrow\rangle\langle\uparrow|)^{\otimes N}\big]$ in the computational basis, the marginal probability of observing an all-down configuration on any subregion $X$ is bounded by $p^{|X|}$, where $|X|$ denotes the size of $X$. An analogous statement holds for $\mathcal{E}\big[(|\downarrow\rangle\langle\downarrow|)^{\otimes N}\big]$.
 
To show that $S_\sigma(D|B)$ decays exponentially, we perform a measurement of all qubits in $B$ in the computational basis, which results in a classical bitstring $Y$. This is a quantum channel acting on $B$ and data processing combined with Fano's inequality implies $S_{\sigma}(D|B) \leq S(D|Y) \leq H_2(P_{\text{decoder}})$, where $D$ is the binary classical label and $Y$ is a classical variable, $H_2(x)=-(1-x)\ln(1-x)-x\ln x$ is the binary entropy, and $P_{\text{decoder}}$ is the decoding error probability for a chosen decoder.
It is well known that for sufficiently small $p$, $P_{\text{decoder}}\leq e^{-|B|/\xi}$ for the majority-vote decoder.
We derive this bound below for completeness.

The majority-vote decoder on $Y$ proceeds by mapping configurations $\{z_j\}$ satisfying $\sum_{j=1}^{N_B} z_j>0$ ($\sum_{j=1}^{N_B} z_j\leq 0$) to the outcome $m=\uparrow$ ($\downarrow$), where $N_B$ denotes the size of region $B$.
For the $d$-dimensional geometry in Fig.~1(a), $N_B\sim l_B l^{d-1}$, and thus exponential decay in $N_B$ implies exponential decay in $l_B$.
The decoding error probability is $P_{\text{decoder}}=[\Pr(m=\downarrow|\alpha=\uparrow)+\Pr(m=\uparrow|\alpha=\downarrow)]/2$. Now, using the $p$-bounded noise assumption, we obtain
\begin{equation}
\Pr(m=\downarrow|d=\uparrow)\leq \sum_{k=N_B/2}^{N_B} \binom{N_B}{k} p^k .
\end{equation}
For $p<1$, one has $p^k\leq p^{N_B/2}$ for all $k\geq N_B/2$, which gives
\begin{equation}
\sum_{k=N_B/2}^{N_B} \binom{N_B}{k} p^k
\leq p^{N_B/2}\sum_{k=N_B/2}^{N_B} \binom{N_B}{k}
\leq p^{N_B/2} 2^{N_B}.
\end{equation}
It follows that $\Pr(m=\downarrow|d=\uparrow)\leq (4p)^{N_B/2}$.
An identical bound holds for $\Pr(m=\uparrow|d=\downarrow)$.
Therefore, for $p<1/4$, the decoding error probability satisfies $P_{\text{decoder}}\leq (4p)^{N_B/2}$ and decays exponentially with $N_B$.
Using the inequality $H_2(x)\leq x[1-\ln x]$, we conclude that $H_2(P_{\text{decoder}})$, and hence $S_\sigma(D|B)$, decays exponentially as a function of $N_B$, and thus as a function of $l_B$. 
 
This discussion generalizes to families of classical $(n,k,d)$ codes whose restriction to $B$ is injective and defines an $(N_B,k,d_B)$ code with $d_B/N_B\geq\delta>0$ uniformly in $N_B$. For example, puncturing $s\leq n/4$ coordinates of a first-order Reed-Muller code with $(n,k,d)=(2^m,m+1,2^{m-1})$ preserves $k$ and gives $N_B=n-s$ and $d_B\geq d-s$, so $d_B/N_B\geq1/3$~\cite{macwilliams1977theory}. For this generalization, we require codeword-uniform $p$-bounded noise: for each transmitted codeword and every $X\subseteq B$, the probability that every bit in $X$ is flipped is at most $p^{|X|}$. We again measure all qubits in $B$, resulting in a classical bitstring $Y$. For a fixed transmitted message $D = i$ (where $i=1,2,\ldots,2^k$), let $c_i$ be the corresponding codeword on $B$ and denote the number of flipped bits in $B$ relative to the transmitted codeword as $|E|$. By the defining property of the code distance $d_B$, if $|E| \leq (d_B-1)/2$, decoding succeeds. Therefore, the decoding error is bounded by $\Pr(|E| \geq d_B/2|D = i) \leq \sum_{j=\lceil d_B/2\rceil}^{N_B} \binom{N_B}{j} p^j$. Using the same series of inequalities as for the repetition code, one finds $\Pr(|E| \geq d_B/2|D = i) \leq 2^{N_B} p^{d_B/2}$. This bound decays exponentially in $N_B$ for any size-independent $p<2^{-2/\delta}$ (note that we recover the bound $p_c = 1/4$ for the repetition code where $d_B = N_B$). For $2^k$ possible messages, Fano's inequality gives $S(D|Y)\leq H_2(P_{\mathrm{decoder}})+P_{\mathrm{decoder}}\ln(2^k-1)$. Since $k\leq N_B$, an exponentially small decoding error therefore implies exponentially small $S(D|Y)$.

\section{Conditional mutual information of classical SSB under decoherence}
\label{sec:cmi_ssb_deco}

The goal of this appendix is to analytically study the CMI of the one-dimensional mixed state
\begin{equation}
\label{Eq:supp_rhopq}
\rho_L(p,q)=\frac{1}{2}\big[(1-q)|\uparrow\rangle\langle\uparrow|+q|\downarrow\rangle\langle\downarrow|\big]^l+\frac{1}{2}\big[(1-p)|\downarrow\rangle\langle\downarrow|+p|\uparrow\rangle\langle\uparrow|\big]^l,
\end{equation}
which can be obtained by subjecting a classical SSB state $\rho_0=\tfrac{1}{2}(|\uparrow\rangle\langle\uparrow|)^l+\tfrac{1}{2}(|\downarrow\rangle\langle\downarrow|)^l$ to a product of single-site channels that flip a $\downarrow$ spin with probability $p$ and an $\uparrow$ spin with probability $q$. A useful property of $\rho_l$ that dramatically simplifies the calculation of the CMI is that the reduced density matrix $\rho_X\equiv\tr_{\bar{X}}\rho_l$ of any region $X$ of size $|X|$ takes the same form as Eq.~\eqref{Eq:supp_rhopq}, namely $\rho_X=\rho_{|X|}=\frac{1}{2}\big[(1-q)|\uparrow\rangle\langle\uparrow|+q|\downarrow\rangle\langle\downarrow|\big]^{|X|}+\frac{1}{2}\big[(1-p)|\downarrow\rangle\langle\downarrow|+p|\uparrow\rangle\langle\uparrow|\big]^{|X|}$. Therefore, the CMI $I(A:C|B)$ of total system size $l = l_B+2r$ and $B$ of size $l_B$ takes the form
\begin{equation}
\label{Eq:supp_ills_cmi}
I(l = l_B + 2r|l_B) = I(r,l_B)=S(AB)+S(BC)-S(B)-S(ABC)=2H_{r+l_B}-H_{l_B}-H_{l_B+2r},
\end{equation}
where $H_r=H_r(p,q)$ denotes the Shannon entropy of the reduced density matrix $\rho_r(p,q)$ and is given by
\begin{equation}
\label{Eq:supp_Hr}
H_r(p,q)=-\sum_{m=0}^r \binom{r}{m} t_{r,m}\log t_{r,m}, \qquad
t_{r,m}=\tfrac{1}{2}\big[(1-q)^{r-m}q^m+p^{r-m}(1-p)^m\big].
\end{equation}
In the following, we show that the CMI becomes singular in the limit $\epsilon=p+q-1\to0^-$, which underlies the phase diagram shown in Fig.~\ref{fig:illustrative_example}. Interestingly, the correlation-length exponent along the $q=0$ line (corresponding to the case studied in the main text) and along the $p=0$ line is $\nu=1$, which differs from the exponent $\nu=2$ obtained when $p,q\neq0$. We note that, unlike the $q=0$ and $p=0$ lines, the case $p,q\neq0$ does not violate the alignment-suppression property discussed in the main text. We first analyze the $q=0$ case and then turn to the case with $p,q\neq0$.

\begin{figure}
\includegraphics[width=0.25\linewidth]{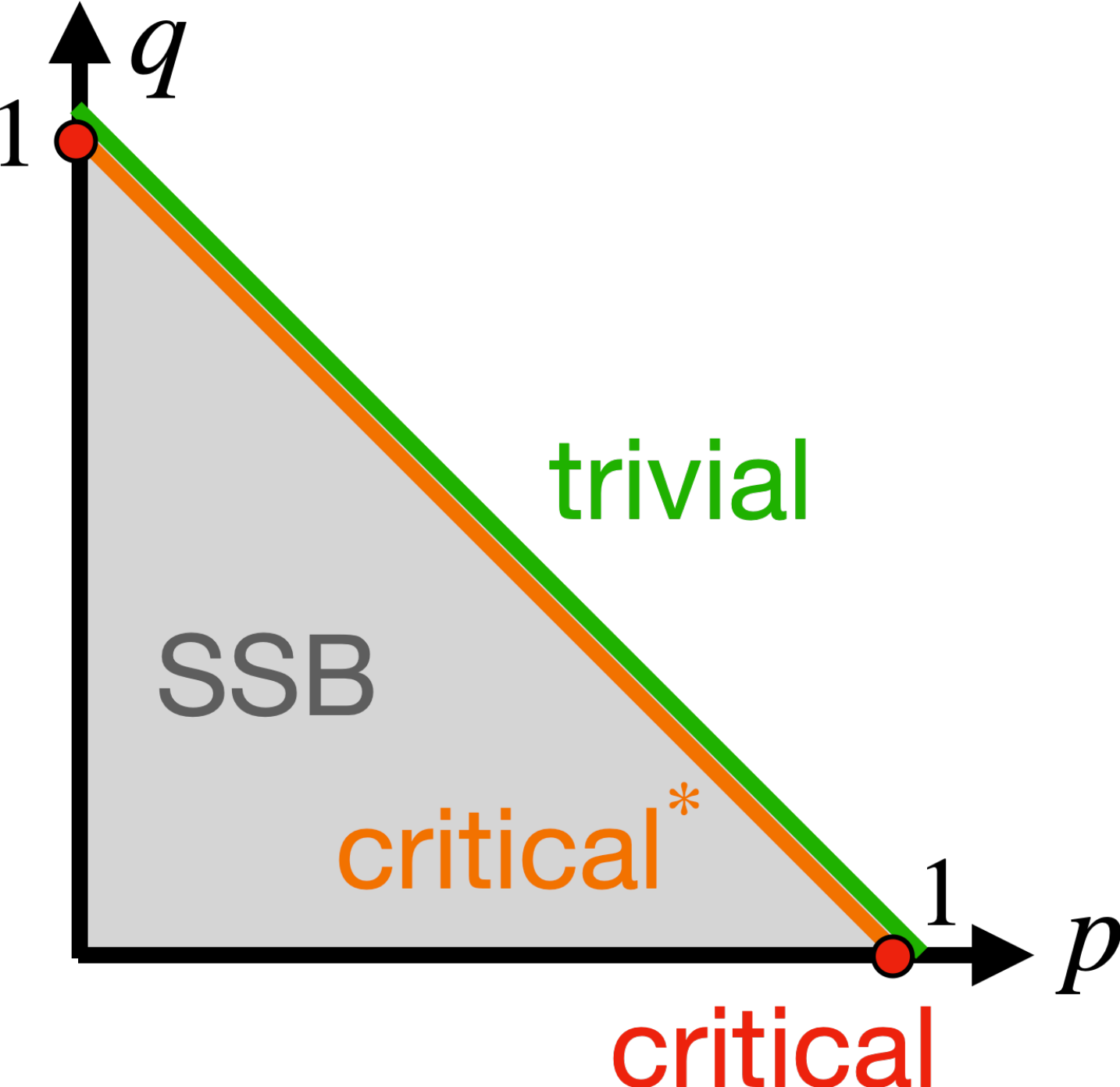}
\caption{ Phase diagram of the mixed state in Eq.\eqref{Eq:supp_rhopq}. The $q = 0$ line corresponds to the case studied in the main text. Interestingly, the correlation-length exponent along the $q=0$ line (corresponding to the case studied in the main text) and along the $p=0$ line is $\nu=1$. This differs from the exponent $\nu=2$ obtained when $p,q\neq0$. We denote the former case as ``critical" while the latter case as ``critical*".
}
\label{fig:illustrative_example}
\end{figure}

\subsection*{CMI along the $q = 0$ line}
Along the $q = 0$ line, the probability  $t_{r,m}$ takes a simpler form
\begin{equation}
    t_{r,m} = \frac{\delta_{m,0} + p^{r-m}(1-p)^m}{2} = 
    \begin{cases}
        \frac{1+p^r}{2},\ m = 0. \\
        \frac{p^{r-m} (1-p)^m}{2}, \ m \geq 1.
    \end{cases}
\end{equation}
It follows that 
\begin{equation}
\begin{aligned}
H_r & = -\frac{1+p^r}{2} \log(\frac{1+p^r}{2}) - \sum_{m =1}^r \binom{r}{m} \frac{p^{r-m} (1-p)^m}{2} \log(\frac{p^{r-m} (1-p)^m}{2}) \\
& = -\frac{1+p^r}{2} \log(\frac{1+p^r}{2}) - \sum_{m =0}^r \binom{r}{m} \frac{p^{r-m} (1-p)^m}{2} \log(\frac{p^{r-m} (1-p)^m}{2}) + \frac{p^r}{2} \log\big(\frac{p^r}{2}\big) \\
& = \log 2 + \frac{r}{2}h(p) -\Big( \frac{1}{2}(1+p^r) \log(1+p^r) - \frac{1}{2}p^r\log(p^r) \Big) \\
& = \log 2 + \frac{r}{2}h(p) -g(r)
\end{aligned}
\end{equation}
where $h(p) = -p\log(p) - (1-p)\log(1-p)$ is the binary entropy and we denote $g(r) = \frac{1}{2}(1+p^r) \log(1+p^r) - \frac{1}{2}p^r\log(p^r) $.  In the third line, we have used the properties that  $\sum_{m =0}^r \binom{r}{m} {p^{r-m} (1-p)^m}  = 1$ and $  \sum_{m =0}^r \binom{r}{m} {p^{r-m} (1-p)^m} \log({p^{r-m} (1-p)^m}) = -rh(p)$. 

Since the constant and linear terms cancel out when computing the CMI [Eq.\eqref{Eq:supp_ills_cmi}], one finds
\begin{equation}
\label{Eq:supp_q0}
I(l_B+2r|l_B,p) = g(l_B,p) +g(l_B+2r,p) - 2g(l_B+r,p).
\end{equation}

\subsubsection*{$r = \infty$}
Let's first consider $r \to \infty$ limit. Since $\lim_{r \to \infty}g(r ,p) =0,\ \forall p <1$, Eq.\eqref{Eq:supp_q0} simplifies to $I(\infty|l_B,p) = g(l_B,p) $. Now, in the limit $l_B \rightarrow \infty$ and $\epsilon = p-1 \rightarrow 0^-$  with $x \equiv l_B \epsilon$ finite, one can write $p^{l_B} = e^{l_B \ln p} \approx e^{l_B \epsilon} = e^{ x}$. It follows that,
\begin{equation}
I(\infty|l_B,p) = g(l_B,p) \longrightarrow g(x) \equiv \frac{1}{2}[(1+e^{x})\log(1+e^{x})-x e^x],
\end{equation}
and thus we've derived Eq.(8) in the main text.

\subsubsection*{$r < \infty$}
As in the previous case, in the limit $l_B \rightarrow \infty$ and $\epsilon = p-1 \rightarrow 0^-$  with $x \equiv l_B \epsilon$ finite, one can write $p^{l_B}= e^{l_B \ln p} \approx e^{l_B\epsilon} = e^{ x}$. Furthermore, since now $r$ is finite, one should treat $\delta = xr/l_B$ as a small parameter in the $l_B \rightarrow \infty$ limit. This implies $p^{l_B+r}= e^{(l_B+r) \ln p} \approx e^{l_B \epsilon (1+r/l_B) } = e^{x+\delta}$, and similarly $p^{l_B+2r} =  e^{x+2\delta}$. It follows that
\begin{equation}
I(r,l_B,p) = g(x) +g(x+2\delta) - 2g(x+\delta ) +\mathcal O\big((r/l_B)^3\big)
= \delta^2 \partial_x^2 g+\mathcal O\big((r/l_B)^3\big)
= \delta^2 \Big(\frac{1}{2} e^{x} \big[ \log(1+e^{-x}) - \frac{1}{1+e^{x}}\big]\Big)+\mathcal O\big((r/l_B)^3\big).
\end{equation}
Substituting $\delta = x r/l_B$, one finds
\begin{equation}
\begin{aligned}
I(r,l_B,p) = \Big(\frac{r}{l_B}\Big)^2   \Big(\frac{x^2}{2} e^{x} \big[ \log(1+e^{-x}) - \frac{1}{1+e^x}\big]\Big) +\mathcal O\big((r/l_B)^3\big)
= \Big(\frac{r}{l_B}\Big)^2 f[(p-1)l_B]+\mathcal O\big((r/l_B)^3\big).
\end{aligned}
\end{equation}
Therefore, one finds the anomalous dimension $\eta = 2$.

\subsection*{CMI for the $p,q \neq 0$ case}

We now consider the CMI for the $p, q\neq 0$ case. Let's first simplify the notation  in Eq.\eqref{Eq:supp_Hr} by denoting $A_m=(1-q)^{r-m}q^{m}$, $
B_m=p^{r-m}(1-p)^{m}$, so that $ 
t_{r,m}=(A_m+B_m)/2 = A_m(1+e^{-r \Delta_m})/2$ with
\begin{equation}
e^{-r \Delta_m}=\frac{B_m}{A_m} = \left(\frac{p}{1-q}\right)^{r-m}\left(\frac{1-p}{q}\right)^{m} = \exp\Big( {-r \big[\big( 1- \frac{m}{r}\big) \log(\frac{1-q}{p}) +\frac{m}{r}\log(\frac{q}{1-p})\big]} \Big)
\end{equation}
Similarly, we write
\begin{equation}
P_q^{(r)}(m)=\binom{r}{m}(1-q)^{r-m}q^{m},\qquad
P_{1-p}^{(r)}(m)=\binom{r}{m}p^{r-m}(1-p)^{m}.
\end{equation}
One can then rewrite Eq.\eqref{Eq:supp_Hr} as
\begin{equation}
\label{eq:HL-decomp}
\begin{aligned}
H_r
&=-\tfrac12\sum_m P_q^{(r)}(m)\Big(\log A_m-\log 2+\log(1+R_m)\Big)
-\tfrac12\sum_m P_{1-p}^{(r)}(m)\Big(\log B_m-\log 2+\log(1+R_m^{-1})\Big)
\\
&=\log 2 + \frac{r}{2}\big[h(q)+h(p)\big]- \frac12\,\mathcal S_{q\to p}(r)- \frac12\,\mathcal S_{p\to q}(r),
\end{aligned}
\end{equation}
where
\begin{equation}
\mathcal S_{q\to p}(r)=\sum_m P^{(r)}_q(m)\big[\log(1+e^{-r \Delta_m})\big],\qquad
\mathcal S_{p\to q}(r)=\sum_m P^{(r)}_{1-p}(m)\big[\log(1+e^{r \Delta_m})\big].
\end{equation}
Since the constant and linear terms cancel out when computing the CMI [Eq.\eqref{Eq:supp_ills_cmi}], one finds
\begin{equation}
I(l = l_B+2r|l_B, p,q) = \frac{1}{2} \Big( [\mathcal S_{q\to p}(l_B)+\mathcal S_{p\to q}(l_B)] + [\mathcal S_{q\to p}(l_B+2r)+\mathcal S_{p\to q}(l_B+2r)] -2 [\mathcal S_{q\to p}(l_B+r)+\mathcal S_{p\to q}(l_B+r)]\Big).
\end{equation}

\subsubsection*{$r = \infty$}
Similar to the $ q = 0$ case, when $r = \infty$, all the terms involving $r$ vanish, and thus one has
\begin{equation}
I(\infty|l_B,p,q) =  \frac{1}{2}  [\mathcal S_{q\to p}(l_B)+\mathcal S_{p\to q}(l_B)]. 
\end{equation}
We now consider the large-$l_B$, small-$|\epsilon|$ limit of $\mathcal S_{q \rightarrow p}(l_B)$, where $\epsilon = p+q-1$. The quantity $\mathcal S_{p \rightarrow q}(l_B)$ is obtained by exchanging $q \leftrightarrow 1-p$ and has the same leading scaling limit as $\mathcal S_{q \rightarrow p}(l_B)$ because $q=1-p+\mathcal O(\epsilon)$. It is more convenient to introduce the variable $u = m/l_B \in[0,1]$ so that
\begin{equation}
\begin{aligned}
\sum_m &\approx l_B \int_{0}^1 du,\quad P^{(l_B)}_q(m) \approx \frac{1}{\sqrt{2 \pi l_B \sigma^2}} e^{- l_B \frac{(u - q)^2}{2 \sigma^2}},\quad \sigma^2 = q(1-q)\\
\Delta_m & = \Delta(u) = \big( 1- u\big) \log(\frac{1-q}{p}) +u\log(\frac{q}{1-p}).
\end{aligned}
\end{equation}
Here we take fixed $0<q<1$ as $l_B\to\infty$, so that $l_B\sigma^2=l_Bq(1-q)\gg1$ and the Gaussian approximation is valid. It follows that
\begin{equation}
\label{Eq:S_large_r_small_e}
\begin{aligned}
\mathcal S_{q\to p}(l_B)& \approx \frac{\sqrt{l_B}}{\sqrt{2\pi \sigma^2}} \int_{0}^{1} du\ e^{-l_B \frac{(u-q)^2}{2\sigma^2}} \log(1+e^{-l_B \Delta(u)}) \\
& = \frac{\sqrt{l_B}}{\sqrt{2\pi \sigma^2}} \int_{-q}^{1-q} du e^{-l_B \frac{u^2}{2\sigma^2}} \log(1+e^{-l_B \Delta(u+q)}) \\
& = \frac{1}{ \sqrt{2\pi}} \int_{-\infty}^{\infty} dv e^{- \frac{v^2}{2}} \log(1+e^{-l_B \Delta\big(\frac{ \sigma v}{\sqrt{l_B}}+q\big)}),
\end{aligned}
\end{equation}
where in the last equality we let $v = \sqrt{l_B} u/\sigma$ so that the end points of the integral go to $\pm \infty$ in the large-$l_B$ limit.
Now, using $p = (1-q) +\epsilon$ and $1-p = q-\epsilon$ with $\epsilon \to 0^-$ and taking $l_B\to\infty$ at fixed $x=\epsilon\sqrt{l_B}/\sigma$, one finds, for fixed $q$ and $v$,
\begin{equation}
\label{Eq:Delta_approx}
l_B \Delta\Big(\frac{ \sigma v}{\sqrt{l_B}}+q\Big) =v \frac{\sqrt{l_B}\epsilon}{\sigma} + \frac{1}{2}\big(\frac{\sqrt{l_B}\epsilon}{\sigma}\big)^2 + \mathcal{O}(l_B^{-1/2})
=vx+\frac{x^2}{2}+\mathcal O(l_B^{-1/2}).
\end{equation}
From Eq.\eqref{Eq:Delta_approx}, it is obvious that the natural scaling variable is ${x} = \epsilon \sqrt{l_B}/\sigma$, and thus the correlation length exponent $\nu = 2$. When $q\to0$ with $q l_B=O(1)$, this Gaussian scaling crosses over to the endpoint regime; at $q=0$, the natural scaling variable is $\epsilon l_B$ and $\nu=1$, as shown above. Therefore, to leading order in the scaling limit,
\begin{equation}
I(\infty|l_B,p,q) \longrightarrow \frac{1}{ \sqrt{2\pi}} \int_{-\infty}^{\infty} dv e^{- \frac{v^2}{2}} \log(1+e^{-\frac{x^2}{2}-v x}) \equiv \Phi(x).
\end{equation}

We note that at the exact endpoint $\epsilon=p+q-1=0$, the two product components in Eq.~\eqref{Eq:supp_rhopq} coincide, so $\rho_L(p,q)$ is a product state and its CMI is zero. By contrast, $\lim_{x\to0^-}\Phi(x)=\log 2$ is the one-sided scaling endpoint obtained after taking the large-$l_B$ scaling limit. Therefore, as in the $q=0$ case discussed in the main text, $\mathrm{UV}\neq\mathrm{UV}^{\prime}$.

\subsubsection*{$r < \infty$}
Similar to the $q=0$ line, in the large-$l_B$, small-$|\epsilon|$ limit, let $h=r/l_B\ll1$. The shifted scaling variables are $\epsilon\sqrt{l_B+r}/\sigma=x[1+h/2-h^2/8+\mathcal O(h^3)]$ and $\epsilon\sqrt{l_B+2r}/\sigma=x[1+h-h^2/2+\mathcal O(h^3)]$. It follows that
\begin{equation}
I(l = l_B + 2r|l_B,p,q)
=h^2\frac{x^2\Phi''(x)-x\Phi'(x)}{4}+\mathcal O(h^3)
=h^2 f(x)+\mathcal O(h^3).
\end{equation}
Therefore, one finds the anomalous dimension $\eta = 2$.

\section{CMI in anisotropic conserved Gaussian dynamics}
\label{sec:supp_soc}

We now provide the technical derivation underlying Illustration 3 in the main
text. 
We first discuss the Gaussian steady-state measure of the anisotropic
conserved dynamics, then isolate the nonlocal terms generated by weak
anisotropy, and finally translate their scaling into the corresponding CMI
scaling. 

\subsection{Steady-state covariance and precision kernel}

We consider the following Langevin equation that describes the anisotropic conserved dynamics
in Ref.~\cite{grinstein1990conservation},
\begin{equation}
\partial_t\phi(\mathbf{x},t)
=
\mu\nabla^2\phi(\mathbf{x},t)
-\nabla^4\phi(\mathbf{x},t)
+\zeta(\mathbf{x},t),
\label{Eq:supp_soc_langevin}
\end{equation}
with $\mu\geq0$. We align the distinguished direction with the
direction normal to the interfaces between $A$, $B$, and $C$. The conserved
Gaussian noise obeys
\begin{align}
\langle\zeta(\mathbf{q},t)\rangle&=0, \nonumber\\
\langle\zeta(\mathbf{q},t)\zeta(\mathbf{q}',t')\rangle
&=
2D(\mathbf{q})\delta(t-t')(2\pi)^d
\delta^d(\mathbf{q}+\mathbf{q}'),
\label{Eq:supp_soc_noise}\\
D(\mathbf{q})
&=
D_{\parallel}q_\parallel^2+D_\perp q_\perp^2,
\nonumber
\end{align}
where $D_{\parallel},D_\perp>0$,
$q_\perp^2\equiv\sum_{\alpha\perp}q_\alpha^2$, and
$q^2=q_\parallel^2+q_\perp^2$. Thus the deterministic relaxation is
rotationally invariant, while the noise distinguishes the direction normal
to the slab from the remaining $d-1$ directions tangent to its interfaces.

For every nonzero momentum, Eq.~\eqref{Eq:supp_soc_langevin} can be written as,
\begin{equation}
\partial_t\phi(\mathbf{q},t)
=
-q^2(\mu+q^2)\phi(\mathbf{q},t)
+\zeta(\mathbf{q},t).
\end{equation}
Its long-time steady-state covariance can then be easily solved as
\begin{equation}
C(\mathbf{q})
\equiv
\langle\phi(\mathbf{q})\phi(-\mathbf{q})\rangle_{\mathrm{ss}}
=
\frac{D_{\parallel}q_\parallel^2+D_\perp q_\perp^2}
{q^2(\mu+q^2)}.
\label{Eq:supp_soc_covariance}
\end{equation}
Writing the steady-state measure as
\begin{equation}
P_{\mathrm{ss}}[\phi]\propto
\exp\left[ 
-\frac{1}{2}
\int_{\mathbf q\neq0} 
\phi(\mathbf q)K(\mathbf q)\phi(-\mathbf q)
\right],
\qquad
\int_{\mathbf q}\equiv
\int\frac{d^d q}{(2\pi)^d},
\end{equation}
we obtain the precision kernel
\begin{equation}
K(\mathbf{q})
=
C(\mathbf{q})^{-1}
=  
\frac{q^2(\mu+q^2)}
{D_{\parallel}q_\parallel^2+D_\perp q_\perp^2}.
\label{Eq:supp_soc_precision}
\end{equation}

Because the dynamics are strictly conserving, neither the relaxation rate nor
the noise fixes the $\mathbf q=0$ mode. In the continuum discussion, we work
on the nonzero-momentum subspace. In the finite-lattice calculation below, we
instead assign the uniform mode a finite covariance. 
We further note that when $D_{\parallel}=D_\perp$, Eq.~\eqref{Eq:supp_soc_precision} reduces to
\begin{equation}
K(\mathbf q)=\frac{\mu+q^2}{D_{\parallel}},
\end{equation}
which is a local differential operator in real space. Its precision block
connecting $A$ directly to $C$ vanishes whenever the two regions are separated
by a nonzero-width buffer $B$. Therefore, the CMI vanishes in the isotropic theory.

\subsection{Weak-anisotropy expansion}

We next expand about the isotropic limit by defining
\begin{equation}
\gamma\equiv\frac{D_\perp}{D_{\parallel}},
\qquad
\delta\equiv1-\gamma,
\qquad |\delta|\ll1.
\end{equation}
We assume $\gamma>0$, so that the only zero of
$q_\parallel^2+\gamma q_\perp^2$ occurs at $\mathbf q=0$. For $\mu>0$, the
crossover length and scaling variable are
\begin{equation}
\xi=\mu^{-1/2},
\qquad
x=\frac{l_B}{\xi}.
\end{equation}
The limit $x\to0^+$ probes distances $l_B\ll\xi$ and approaches the
$\mu=0$ UV fixed point, whereas $x\to\infty$ probes the IR regime in which
the $q^4$ term in the dynamics is irrelevant.

It is illuminating to separate the exact precision kernel into local
and nonlocal parts:
\begin{equation}
\begin{aligned}
K(\mathbf q)
&=K_{\mathrm{loc}}(\mathbf q)
+\frac{1}{D_{\parallel}}
\frac{\mu\delta q_\perp^2+\delta^2q_\perp^4}
{q_\parallel^2+(1-\delta)q_\perp^2},\\
K_{\mathrm{loc}}(\mathbf q)
&=\frac{1}{D_{\parallel}}
\left[\mu+q_\parallel^2+(1+\delta)q_\perp^2\right].
\end{aligned}
\label{Eq:supp_soc_precision_decomposition}
\end{equation}
Because $K_{\mathrm{loc}}$ is polynomial in momentum, it contains only local
contact terms. Therefore, the nonlocal kernel entering CMI starts at $O(\delta^2)$ in the
UV, where $\mu=0$, and at $O(\delta)$ in the IR, where the $q^4$ term
is irrelevant.

\raggedbottom 

\subsection{Direct thermodynamic-limit calculation in two dimensions}

We now explicitly derive the asymptotic form of the area-law coefficient of CMI in
$d=2$. 
Specifically, writing $\text{CMI} = (1/y)g(x,y;\delta)$, we will show that $g_0(x,\delta) \equiv \lim_{y \to 0^+} g(x,y;\delta) =\delta^2 x/(120\pi) + o(x)$ when $\delta^2\ll x\ll 1$.
We take $A$ and $C$ to be semi-infinite
in the normal direction and Fourier transform parallel to their interfaces.
For the Gaussian field conditioned on $B$, let $K_A$ and $K_C$ be the diagonal
precision blocks and $K_{A,C}$ the block coupling the two sides. The exact CMI is
\begin{equation}
I(A:C|B)
=-\frac12\operatorname{Tr}\log
\left(1-\mathcal M\mathcal M^\dagger\right),
\qquad 
\mathcal M=K_A^{-1/2}K_{A,C}K_C^{-1/2}.
\label{Eq:supp_soc_gaussian_cmi}
\end{equation}
Decomposing $\mathcal M$ into transverse-momentum sectors, its restriction to
sector $q$ is
$\mathcal M_q=(K_A^{(q)})^{-1/2}K_{A,C}^{(q)}\allowbreak
(K_C^{(q)})^{-1/2}$.
Let $s_{q,n}$ be its singular values. This sector contributes
\begin{equation} 
I_q=-\frac12\sum_n\log(1-s_{q,n}^2)
=\frac12\sum_n s_{q,n}^2+O\!\left(\sum_n s_{q,n}^4\right).
\label{Eq:supp_soc_gaussian_cmi_singular_values}
\end{equation}
By the Schur complement, positivity implies
$1-\mathcal M_q\mathcal M_q^\dagger>0$. Hence $s_{q,n}<1$ and the first equality
is exact; the expansion further requires $s_{q,n}\ll1$, as we verify uniformly
in $q$ below.

We now determine the ingredients needed to compute $I_q$ through $O(\delta^2)$.
At $\delta=0$, $K_{A,C}^{(q)}=\mathcal M_q=0$. Thus, the $O(\delta^2)$ CMI
requires $K_{A,C}^{(q)}$ to $O(\delta)$, but only $K_A^{(q)}$ and $K_C^{(q)}$
at $\delta=0$; their $O(\delta)$ corrections affect $\mathcal M_q$ only at
$O(\delta^2)$.
The common factor $D_\parallel^{-1}$ in 
Eq.~\eqref{Eq:supp_soc_precision_decomposition} cancels from $\mathcal M_q$ and
will be omitted.

Let's first compute $K_{A,C}^{(q)}$ to $O(\delta)$. For $q>0$, let $u,v\geq0$
measure the distances into $A$ and $C$ from their respective interfaces with
$B$.
Fourier transforming the $O(\delta)$ term in
Eq.~\eqref{Eq:supp_soc_precision_decomposition} along the normal direction
gives
\begin{equation}
\left.\partial_\delta K_{A,C}^{(q)}(u,v)\right|_{\delta=0}
=\int_{-\infty}^{\infty}\frac{dk}{2\pi}\,
\frac{\mu q^2e^{ik(l_B+u+v)}}{k^2+q^2}
=\frac{\mu q}{2}e^{-q(l_B+u+v)}.
\label{Eq:supp_soc_cross_derivative}
\end{equation}
With $a_q=(\mu q/2)e^{-ql_B}$ and $\langle u|f_q\rangle=e^{-qu}$,
Eq.~\eqref{Eq:supp_soc_cross_derivative} is the rank-one operator
$\left.\partial_\delta K_{A,C}^{(q)}\right|_{\delta=0}
=a_q|f_q\rangle\langle f_q|$.

We next evaluate $K_A^{(q)}$ and $K_C^{(q)}$ at isotropy.
Conditioning on $B$ fixes the boundary fluctuations, giving
$K_A^{(q)}=K_C^{(q)}=L_q=-\partial_u^2+q^2+\mu$ on the half-line with a
Dirichlet boundary. Its inverse acts on $f_q$ as
\begin{align}
L_q^{-1}f_q(u)
&=\frac{e^{-qu}-e^{-\sqrt{q^2+\mu}\,u}}{\mu},\nonumber\\
\langle f_q,L_q^{-1}f_q\rangle
&=\frac{1}{2q\left(q+\sqrt{q^2+\mu}\right)^2}.
\label{Eq:supp_soc_halfline_response}
\end{align}
It follows that
\begin{equation}
\mathcal M_q
=\delta a_q|L_q^{-1/2}f_q\rangle
\langle L_q^{-1/2}f_q|+O(\delta^2).
\end{equation}
The leading term is rank one and therefore has a single nonzero singular
value. Using
Eq.~\eqref{Eq:supp_soc_halfline_response}, we obtain
\begin{equation}
s_q=|\delta|a_q\langle f_q,L_q^{-1}f_q\rangle+O(\delta^2)
=\frac{|\delta|\mu e^{-ql_B}}
{4\left(q+\sqrt{q^2+\mu}\right)^2}
+O(\delta^2).
\end{equation}
Since $e^{-ql_B}\leq1$ and
$q+\sqrt{q^2+\mu}\geq\sqrt\mu$, we have
$s_q\leq|\delta|/4+O(\delta^2)$ uniformly for $q>0$. Thus weak anisotropy
controls the expansion in Eq.~\eqref{Eq:supp_soc_gaussian_cmi_singular_values},
even as $q\to0^+$.
Now, substituting this result into
Eq.~\eqref{Eq:supp_soc_gaussian_cmi_singular_values}, using
$g_0=l_B I/L_\perp$ in $d=2$, and setting $q=\sqrt\mu\sinh t$ after summing
the transverse modes yields
\begin{align}
\left.\frac12\partial_\delta^2g_0(x;\delta)\right|_{\delta=0}
&=l_B\int_{-\infty}^{\infty}\frac{dq}{2\pi}
\frac{\mu^2e^{-2|q|l_B}}
{32\left(|q|+\sqrt{q^2+\mu}\right)^4}\nonumber\\
&=\frac{x}{32\pi}\int_0^\infty dt\,
\cosh t\,e^{-4t-2x\sinh t}.
\label{Eq:supp_soc_exact_derivative_continuum}
\end{align}
We note that the isolated conserved mode at $q=0$ has zero measure in this integral and
does not affect the result.
Using
\begin{equation}
\int_0^\infty dt\,\cosh t\,e^{-4t}
=\frac4{15},
\end{equation}
One find that 
\begin{equation}
\left.\frac12\partial_\delta^2g_0(x;\delta)\right|_{\delta=0}
=
\frac{x}{120\pi}+o(x)
\qquad (x\to0^+),
\label{Eq:supp_soc_exact_small_x}
\end{equation}
which is the desired result.

We next check the result at finite $\delta$ by evaluating
Eq.~\eqref{Eq:supp_soc_gaussian_cmi} with the full precision blocks for each
transverse mode and then integrating over momentum. In particular, we compute $[g_0(x;\delta)+g_0(x;-\delta)-2g_0(x;0)]/(2\delta^2)$, which is anticipated to be equivalent to $\left.\frac12\partial_\delta^2g_0(x;\delta)\right|_{\delta=0}$ as $|\delta| \to 0$.
Figure~\ref{fig:supp_anisotropic_cmi_numerics} shows the normalized derivative.
We observe that at any fixed $x$, the finite-$\delta$ results converge to this
curve as $|\delta|$ decreases (their departure at $x\sim\delta^2$ reflects the
nonuniform crossover to the $O(\delta^4)$ UV fixed point). 

\begin{figure}[H]
\centering
\includegraphics[width=0.6\linewidth]{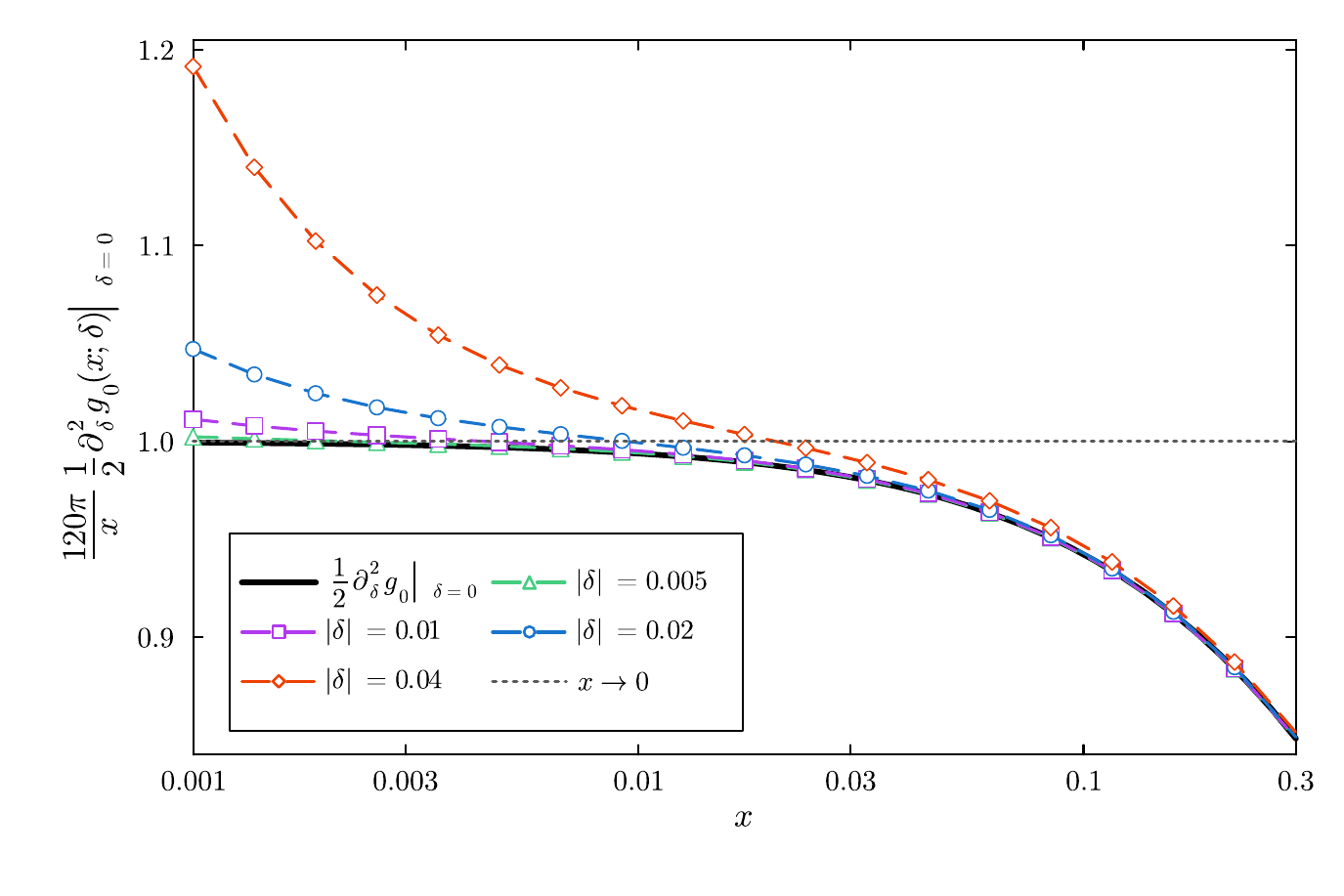}
\caption{\textbf{Thermodynamic-limit verification of the
$O(\delta^2x)$ contribution in $d=2$.}
The black curve shows
$\frac{120\pi}{x}\frac12
\left.\partial_\delta^2g_0(x;\delta)\right|_{\delta=0}$
evaluated from Eq.~\eqref{Eq:supp_soc_exact_derivative_continuum}; its approach
to the dotted line at unity verifies Eq.~\eqref{Eq:supp_soc_exact_small_x}.
Open symbols show the finite-anisotropy symmetric quotient
$\frac{120\pi}{x}
[g_0(x;\delta)+g_0(x;-\delta)-2g_0(x;0)]/(2\delta^2)$,
obtained by evaluating Eq.~\eqref{Eq:supp_soc_gaussian_cmi} with the full
finite-$\delta$ precision kernel at
$|\delta|=0.005$, $0.01$, $0.02$, and $0.04$. At fixed $x$, these data
approach the black curve as $|\delta|$ decreases.}
\label{fig:supp_anisotropic_cmi_numerics}
\end{figure}
\flushbottom

\section{A brief introduction to SWSSB}

The goal of this appendix is to provide a brief review of the strong-to-weak spontaneously symmetry-breaking (SWSSB) phase (See Refs.\cite{lee2023quantum, lessa2024strong} for details). A density matrix is said to have a \textit{strong symmetry} if $U\rho=e^{i\theta}\rho$, and a \textit{weak symmetry} if $U\rho U^{\dagger}=\rho$, where $U$ is the generator of the symmetry \cite{de2022symmetry,ma2022average}. Physically, a strong symmetry is a property of individual states, whereas a weak symmetry holds only at the level of the ensemble in general.

For the application in the main text, the relevant symmetry is generated by $U=\prod_{j=1}^N X_j$ where $N$ is the total system size. In this setting, a mixed state $\rho$ is said to be an SWSSB state if the following three conditions are simultaneously satisfied \cite{lessa2024strong}:
\begin{enumerate}
\item it is strongly symmetric;
    \item there is no long-range order in the two-point correlation function, i.e.,
    $\lim_{|i-j|\to\infty}\tr(\rho\,Z_i Z_j)=0$;
    \item the fidelity between $\rho$ and $Z_i Z_j\rho Z_i Z_j$ saturates to a finite constant as $|i-j|\to\infty$, i.e.,
    $\lim_{|i-j|\to\infty}F(\rho,Z_i Z_j\rho Z_i Z_j)=c>0$, where
    $F(\rho,\sigma)=\tr\!\left(\sqrt{\sqrt{\rho}\,\sigma\,\sqrt{\rho}}\right)$.
\end{enumerate}

Either of the first or the second condition excludes the weakly symmetric, classical SSB state $\rho_{\textrm{classical}} = \tfrac{1}{2} (|\uparrow\rangle\langle\uparrow|)^N + \tfrac{1}{2}(|\downarrow\rangle\langle\downarrow|)^N$, while the second condition excludes the strongly symmetric, quantum SSB fixed point
$\rho_{\text{GHZ}}=|\text{GHZ}\rangle\langle\text{GHZ}|$, where
$|\text{GHZ}\rangle=(|\uparrow\cdots\uparrow\rangle+|\downarrow\cdots\downarrow\rangle)/\sqrt{2}$. The state $\rho_{\textrm{classical}}$ exhibits nonvanishing two-point correlations and a non-trivial fidelity correlator, and therefore breaks the weak symmetry to nothing. On the other hand,
the state $\rho_{\text{GHZ}}$ also exhibits both nonvanishing two-point correlations and a nontrivial fidelity correlator, and can therefore be viewed as spontaneously breaking the strong symmetry to nothing.

By contrast, the SWSSB fixed point takes the form
$\rho_{\text{SWSSB}}=(I+U)/2^{N}$, which can be interpreted as a maximally mixed state projected onto the global $\mathbb{Z}_2$ charge-even sector. Since
$Z_i Z_j\rho_{\text{SWSSB}}Z_i Z_j=\rho_{\text{SWSSB}}$ for all $i,j$, one has
$F(\rho_{\text{SWSSB}},Z_i Z_j\rho_{\text{SWSSB}}Z_i Z_j)=1$.
We further note that $\rho_{\text{SWSSB}}$ has CMI $\log 2$ for a three-region partition in which $A\cup B\cup C$ covers the entire system and each region is nonempty, since the entropy of any proper subsystem $X$ of size $|X|$ is $S_X=|X|\log 2$, while the entropy of the full system is $S=(N-1)\log 2$. For a proper local region $A\cup B\cup C$ embedded in a larger system, tracing out the complement removes the global parity constraint, and the CMI of $\rho_{\mathrm{SWSSB}}$ vanishes. Physically, this nontrivial CMI originates from the global constraint and local irrecoverability \cite{lessa2024strong}.

A well-known transition between a trivial state and the SWSSB state is realized by subjecting a symmetric product state $(|+\rangle\langle+|)^{L_x L_y}$ on a square lattice to the channel
$\mathcal{E}_{\langle i,j\rangle}[\cdot]=(1-p)(\cdot)+p\,Z_i Z_j(\cdot)Z_i Z_j$
acting on all nearest-neighbor pairs \cite{lee2023quantum}. In this setting, the mixed state is always diagonal in the Pauli-$X$ basis and can be written as
$\rho(p)\propto\sum_{x_{\mathbf{j}}\ \text{s.t.}\ \prod_j x_j=1}\mathcal{Z}_{x_{\mathbf{j}}}(p)\,|x_{\mathbf{j}}\rangle\langle x_{\mathbf{j}}|$,
where $|x_{\mathbf{j}}\rangle = |x_1, \cdots, x_{L_x L_y}\rangle,\ x_j = \pm 1$ denotes a product state in the Pauli-$X$ basis. The weight
$\mathcal{Z}_{x_{\mathbf{j}}}(p)=\sum_{\{s_{\tilde{i}}\}} e^{\beta\sum_{\langle\tilde{i},\tilde{j}\rangle} J_{\langle\tilde{i},\tilde{j}\rangle} s_{\tilde{i}} s_{\tilde{j}}}$,
with $\tanh\beta=1-2p$, is the partition function of the random-bond Ising model defined on the dual lattice $\{\tilde{i}\}$. Here the bond configuration $\{J_{\langle\tilde{i},\tilde{j}\rangle}\}$ is constrained to satisfy
$\prod_{\langle\tilde{i},\tilde{j}\rangle\in j} J_{\langle\tilde{i},\tilde{j}\rangle}=x_j$. We emphasize that $\rho(p)$ is generally not a Gibbs state of a local Hamiltonian, due to the summation $\sum_{\{s_{\tilde{i}}\}}$ appearing in the weight $\mathcal{Z}_{x_{\mathbf{j}}}(p)$.
A straightforward calculation shows that the fidelity correlator is related to the free-energy cost of the random-bond Ising model along the Nishimori line, which is nonvanishing only for $p>p_c\approx 0.109$ \cite{nishimori1981internal}. Consequently, the universality class of this transition is that of the random-bond Ising model's Nishimori multicriticality.

\section{Additional data for the CMI of SWSSB}

The goal of this appendix is to provide numerical evidence for UV-finite CMI across the SWSSB transition. To minimize finite-size effects arising from additional length scales, we consider the geometry shown in Fig.~\ref{fig:SWSSB_cmi_scale}(a), where regions $A$, $B$, and $C$ are all $l\times l$ boxes. Fig.~\ref{fig:SWSSB_cmi_scale}(b) shows the CMI as a function of $p$ for $l=12,16,20,24,$ and $28$. We use $p=0.01,0.02,\ldots,0.21$, including $p=0.10,0.11,0.12,$ and $0.13$ near $p_c$, and average over $1000$ independent samples at each point; the error bars show one standard error of the mean. Fig.~\ref{fig:SWSSB_cmi_scale}(c) shows a data collapse using the scaling ansatz $I=f[(p-p_c)l^{1/\nu}]$ with $p_c\approx 0.109$ and $\nu=1.5$. 
The observed collapse provides numerical evidence for UV-finite CMI.

\begin{figure}
\centering
\includegraphics[width=\linewidth]{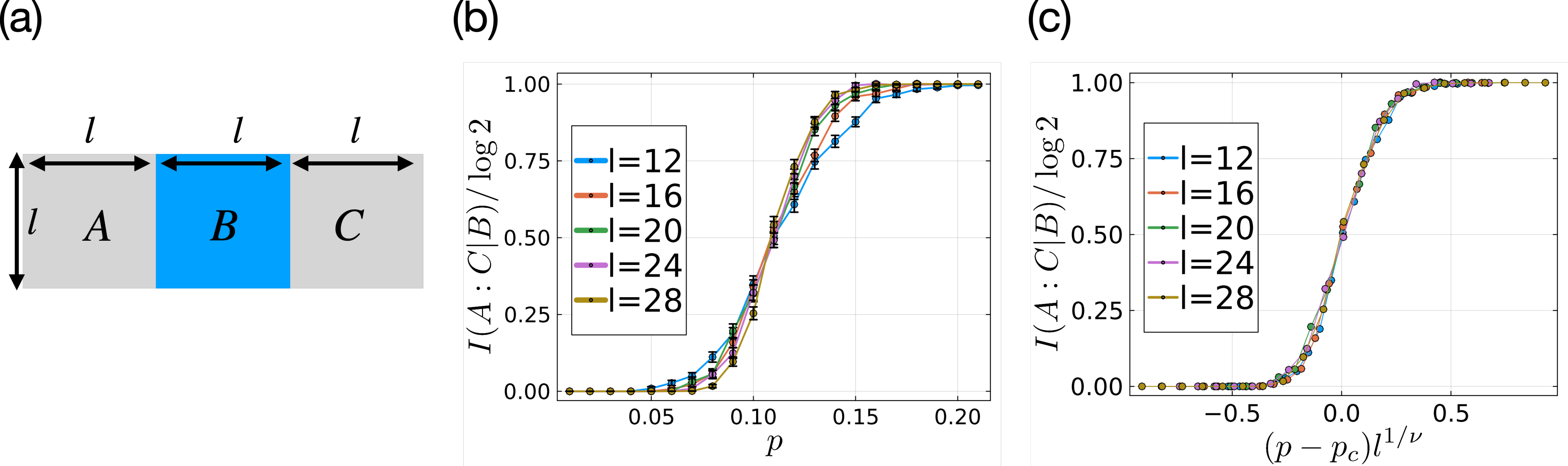}
\caption{(a) Geometry used to compute the CMI across the trivial-to-SWSSB transition. (b) CMI as a function of $p$ for various values of $l$. (c) Data collapse using the scaling ansatz $I=f[(p-p_c)l^{1/\nu}]$ with $p_c=0.109$ and $\nu=1.5$.
}
\label{fig:SWSSB_cmi_scale}
\end{figure}

\end{document}